\documentclass[11pt]{article}

\usepackage[utf8]{inputenc}
\usepackage[T1]{fontenc}
\usepackage[english]{babel}

\usepackage{amsmath,amssymb,amsfonts}
\usepackage{graphicx,booktabs,array,longtable,multirow,threeparttable}
\usepackage{caption,subcaption}
\usepackage{float}
\usepackage{enumitem}
\usepackage{geometry}
\usepackage{natbib}
\usepackage[section]{placeins}

\usepackage[hidelinks]{hyperref}

\usepackage[hidelinks]{hyperref}
\hypersetup{
    colorlinks=true,
    linkcolor=blue,
    citecolor=blue,
    urlcolor=blue
}

\usepackage{xcolor}
\usepackage{float} 
\usepackage{footmisc} 
\begin{document}

\begin{titlepage}
    \centering
    \vspace*{3cm}
    {\LARGE\textbf{Words Matter: Forecasting Economic Downside Risks with Corporate Textual Data}}\\[1cm]
    {\large Cansu Isler\footnote{Ph.D Candidate, Brandeis University. Contact: cansuisler@brandeis.edu. I am deeply grateful to my advisors, Davide Pettenuzzo and Yeabin Moon, for their invaluable guidance and continuous support. I also thank Catherine L. Mann, Giulia Mantoan, and seminar participants at Brandeis University for their insightful comments and stimulating discussions.}}\\[0.5cm]
    {\large \today}\\[2cm]

\begin{abstract}
\noindent
Accurate forecasting of downside risks to economic growth has become critically important for policymakers and financial institutions, particularly following recent economic crises. This paper builds upon the Growth-at-Risk (GaR) approach, introducing a novel daily sentiment indicator derived from textual analysis of mandatory corporate disclosures (SEC 10-K and 10-Q reports) to forecast downside risks to economic growth. Utilizing the Loughran–McDonald dictionary and word-count methodology, I calculate firm-level tone growth by measuring the year-over-year difference between positive and  negative sentiment expressed in corporate filings. These firm-specific sentiment metrics are aggregated into a weekly tone index, weighted by firms' market capitalizations to capture broader, economy-wide sentiment dynamics. Integrated into a Mixed Data Sampling (MIDAS) Quantile regression framework, this sentiment-based indicator enhances the prediction of GDP growth downturns, outperforming traditional financial market indicators such as the National Financial Conditions Index (NFCI). The findings underscore corporate textual data as a powerful and timely resource for macroeconomic risk assessment and informed policymaking.

\vspace{0.5cm}

\end{abstract}

\end{titlepage}
\section{Introduction} 
Accurately forecasting downside risks to economic growth is a central concern for policymakers and financial institutions. This concern has intensified in the wake of severe downturns such as the 2008 global financial crisis and the 2020 COVID-19 shock. Traditional macroeconomic forecasts, which often focus on the mean or baseline outlook, have been criticized for failing to anticipate such tail-risk events. In response, a growing literature has shifted toward modeling the entire distribution of future GDP growth rather than just point forecasts. In particular, the ``Growth-at-Risk'' (GaR) framework links current financial and economic conditions to the conditional distribution of future growth, providing estimates of the probability of adverse outcomes in GDP growth \citep{Adrian2019, Prasad2019}. This approach is analogous to the concept of Value-at-Risk in finance but applied to aggregate economic activity: it quantifies how vulnerabilities in the present can translate into higher likelihood of low (or negative) growth in the future. Empirical studies find that downside risks to GDP growth vary over time and are heavily influenced by prevailing financial conditions \citep{Adrian2019}. Notably, the left tail of the future growth distribution tends to deteriorate markedly when financial markets are under stress, even if the central forecast remains stable. This ability to assess the entire forecast distribution (and especially its lower tail) makes GaR a powerful tool for macro-financial surveillance and has been adopted by institutions like the International Monetary Fund for risk monitoring \citep{Prasad2019}.\\~\\
However, existing GaR models and related macroeconomic risk forecasts often rely on indicators that may be overly backward-looking or slow-moving. Many studies use aggregate financial conditions indices—such as the National Financial Conditions Index (NFCI)—to proxy for macro-financial conditions. While these indices capture contemporaneous financial tightness or stress, they largely reflect information from recent historical data and traditional economic indicators. As a result, their predictive content for future downturns can be limited. For example, \citet{Hatzius2010} find that several popular financial conditions indexes have little out-of-sample forecasting advantage for growth relative to simpler predictors, except during periods around the crisis they were designed to track. Similarly, \citet{Kliesen2012} document that many financial stress indices are highly persistent and tied to past crisis episodes, raising concerns about their utility in signaling emerging risks.\\~\\
In this paper, I propose a text-based approach to enhance forecasts of downside macroeconomic risk by extracting forward-looking sentiment from corporate disclosures. Firms' mandatory reports to investors—especially annual 10-K and quarterly 10-Q filings submitted to the U.S. Securities and Exchange Commission (SEC)—contain extensive qualitative information about business prospects and risks. Senior executives discuss financial results alongside risk factors, uncertainties, and outlooks, making these disclosures timely indicators of corporate sentiment. An increasingly pessimistic tone or heightened discussions of risk across many firms may signal broader impending economic slowdowns. My approach builds on the burgeoning literature in financial textual analysis, which demonstrates that the content and tone of corporate documents can meaningfully predict economic outcomes \citet{Loughran2011, Hassan2019}.\\~\\
Specifically, I construct a new firm-level sentiment indicator using the textual tone of 10-K and 10-Q reports, applying finance-specific sentiment dictionaries and natural language processing techniques \citep{Loughran2011}. The resulting high-frequency (weekly) sentiment indicator aggregates sentiment across numerous corporate filings, capturing management’s private forward-looking information regarding demand conditions, investment plans, and perceived risks before they manifest in traditional macroeconomic indicators. This study complements recent research emphasizing the predictive power of textual analysis, including news-based indexes and corporate reports \citep{Baker2016,Hassan2023,Bybee2021}.\\~\\
A critical methodological challenge in using textual sentiment to forecast macroeconomic outcomes is the mismatch between high-frequency sentiment measures (weekly or daily) and low-frequency GDP data (quarterly). I bridge this gap through Mixed Data Sampling (MIDAS) quantile regression methods introduced by \citet{Ghysels2007}. MIDAS regression efficiently incorporates high-frequency predictors into low-frequency regression models without requiring temporal aggregation. I further adapt MIDAS to a quantile regression framework, emphasizing tail-risk forecasting \citep{Pruser2024, Andreou2013}. This approach allows immediate utilization of changes in corporate sentiment and recognizes potentially nonlinear impacts of sentiment on different quantiles of GDP growth, particularly severe downturns.\\~\\
This paper makes three key contributions:
\begin{enumerate}
\item \textit{Novel sentiment-based risk indicator:} I develop a new forward-looking sentiment indicator derived from corporate disclosures, providing timely signals of macroeconomic vulnerabilities not fully captured by traditional backward-looking financial indicators.
\item \textit{Methodological innovation – MIDAS quantile regression:} I integrate high-frequency corporate textual sentiment into a quarterly macroeconomic forecasting framework, demonstrating significant advantages over traditional temporal aggregation methods.

\item \textit{Practical policy implications:} Empirical results indicate that my sentiment-based indicator significantly enhances forecasts of GDP downside risks, offering policymakers an improved tool for early detection and management of economic vulnerabilities.
\end{enumerate}
Overall, my study bridges macro-finance literature on Growth-at-Risk and financial textual analysis literature, providing novel evidence that forward-looking corporate narratives significantly inform macroeconomic risk assessments.\\~\\
The rest of the paper is organized as follows. Section 2 reviews the related literature. Section 3 describes the data used in the analysis. Section 4 details the construction of the corporate sentiment indicator and outlines the MIDAS quantile regression methodology. Section 5 presents the empirical results, including both in-sample analysis and out-of-sample forecasting performance, and it provides robustness checks. Section 6 concludes.
\section{Literature Review} 
\subsection{Textual Sentiment Analysis in Finance and Macroeconomics} 
"Over the past two decades, researchers have increasingly used textual analysis to extract sentiment and predictive signals from unstructured data in finance and macroeconomics. Early seminal work demonstrated the value of qualitative text for understanding market behavior. For example, \citet{Tetlock2007} quantitatively analyzed a popular Wall Street Journal column and found that high pessimistic tone in media coverage predicts short-term stock price declines followed by reversion to fundamentals. This study gave concrete evidence that investor sentiment, as reflected in news, can influence asset prices. Subsequent research in finance solidified the importance of textual sentiment measures. In particular, \citet{Loughran2011} developed a domain-specific sentiment dictionary tailored to financial language, showing that general-purpose sentiment lexicons (e.g. Harvard IV) misclassify common finance terms (such as “liability” or “capital”) as negative, whereas a finance-specific dictionary yields more meaningful tone measures for corporate disclosures. Using their word lists on firms’ 10-K reports, they demonstrated that a negative tone in annual reports is associated with lower future earnings and stock returns, highlighting that qualitative tone in mandatory filings provides information beyond quantitative accounting data. In the macroeconomic realm, textual analysis has been used to gauge broad economic sentiment and uncertainty. \citet{Baker2016}, for instance, introduced the Economic Policy Uncertainty index by counting policy-related keywords in newspapers, and showed that spikes in this text-derived uncertainty measure foreshadow declines in investment, output, and employment. This work, published in a top economics journal, underscored that newspaper text can be systematically quantified to capture policy-driven economic sentiment with real effects. More recently, \citet{Husted2020} construct a Monetary Policy Uncertainty index from Federal Reserve documents, illustrating how textual cues from structured central bank communications can be quantified to reflect uncertainty around policy actions. Overall, these studies established that analyzing text—from news media, firm disclosures, or central bank communications—yields sentiment metrics that often have significant predictive power for financial markets and the macroeconomy.\\~\\
Methodologically, approaches to textual sentiment analysis in finance and economics can be broadly divided into dictionary-based and machine-learning-based techniques. Dictionary (or lexicon) methods remain popular for their simplicity. In this approach, one defines a lexicon of positive, negative, and uncertainty words and measures sentiment by counting their occurrences in a document (possibly scaled by total words or adjusted for context). Influential early studies like \citet{Tetlock2007} and \citet{Loughran2011} employed dictionary-based sentiment – the former using a general sentiment word list (Harvard’s negative word list) and the latter creating a finance-specific dictionary – to successfully link textual tone with economic outcomes. The appeal of dictionary methods lies in their interpretability: researchers can easily understand which words drive the sentiment score, making it straightforward to interpret “what the text is saying.” However, a well-known limitation is that fixed dictionaries may fail to capture contextual nuances or evolving language usage. For example, a dictionary cannot on its own distinguish negation (“not good”) or sarcasm, nor can it adapt when new jargon or firm-specific language emerges. These limitations have motivated the use of more complex machine learning (ML) and natural language processing (NLP) techniques. ML-based approaches treat sentiment analysis as a predictive classification or regression problem, using statistical models trained on example texts to infer sentiment. Such models can be supervised (trained on texts labeled with sentiment or with an outcome of interest) or unsupervised (learning latent topics or themes from text). In economics, \citet{Gentzkow2019} survey a wide array of text-as-data methods, highlighting how techniques like topic modeling, clustering, and tree-based classifications have been applied to extract information from text beyond simple word counts. For instance, \citet{JegadeeshWu2013} present a novel data-driven content analysis method: rather than relying on an external dictionary, they infer the tone of corporate reports by statistically learning which words in 10-K filings predict firms’ subsequent stock returns. Their algorithm assigns weights to words based on market reactions, effectively creating a customized sentiment measure with improved predictive power for returns. This approach is akin to a machine-learning model that “trains” on historical text and outcome data to identify which patterns of language are truly meaningful. The advantage of such ML approaches is greater flexibility and potentially higher out-of-sample accuracy – they can capture subtle linguistic cues (tone, context, even syntax) that fixed dictionaries might misclassify. Indeed, applications of deep learning and transformer-based models (fine-tuned for financial text) have shown further improvements in capturing the nuance of sentiment in news and social media. The downside is that these models often act as “black boxes,” making it harder to interpret what drives their sentiment scores; they also require large training corpora and careful tuning to avoid overfitting noise. In practice, both dictionary and ML approaches are viewed as complementary. Many finance studies start with dictionary-based sentiment as a baseline, given its ease of use, and then explore ML enhancements for additional predictive gain. As \citet{Gentzkow2019} note, the choice often depends on the context and available data: if human labels or clear objective outcomes are available to train on, supervised ML can outperform lexicon methods, whereas in data-scarce settings a well-crafted dictionary may suffice and is less prone to spurious patterns.\\~\\
Another important dimension in this literature is the nature of the textual sources: structured vs. unstructured disclosures, and firm-level vs. media-based text data. Structured disclosures refer to texts that follow a regulated or standardized format, such as corporate filings (annual reports, quarterly reports) or transcripts of earnings conference calls, which tend to have formal language and recurring sections (e.g. discussions of financial performance, risks, outlook). Unstructured texts include news articles, social media posts, or other free-form communications that are not standardized in presentation or content. Each type of source offers distinct advantages and challenges for sentiment analysis. Firm-level disclosures, structured texts produced directly by companies, often contain rich insider information about business conditions and expectations. Managers may reveal optimism or concern about future earnings, project pipelines, or macro conditions in the narrative sections of reports or calls. These texts are usually carefully worded, yet researchers have found they still contain predictive signals. For example, the tone in annual reports and IPO prospectuses has been shown to predict firm performance and risk \citep{Loughran2011, JegadeeshWu2013}. \citet{Hassan2019} demonstrate that by aggregating firm-level textual information one can measure economy-wide risks: they analyze thousands of quarterly earnings call transcripts to construct a novel index of firm-level political risk, then show that elevated political risk in these transcripts correlates with reduced corporate investment and hiring at the macro level. This underscores a benefit of firm-level textual data: micro-level sentiments can be aggregated to glean bottom-up insights about the broader economy’s state (e.g. if many firms lament weak demand or policy uncertainty, it likely signals macroeconomic headwinds). However, a challenge with firm disclosures is their lower frequency and coverage – e.g. 10-K reports are annual, and even quarterly calls may be too infrequent to capture rapidly changing conditions. Moreover, private firms or small enterprises (which collectively matter for the economy) do not produce similar publicly available texts, so macro sentiment indices built from firm communications might be biased toward larger public companies.\\~\\
In contrast, media-based textual data – such as newspapers, financial news wires, and other journalistic sources – provide higher-frequency and broad coverage of information, albeit in a less structured form. Media outlets continuously interpret and synthesize diverse signals (firm news, economic data releases, policy changes, geopolitical events), so the sentiment in media can serve as a near real-time barometer of public and investor mood. \citet{Tetlock2007}’s media sentiment measure is an early example focusing on a daily news column; subsequent studies have extended this to a variety of outlets and time periods. Notably, \citet{Manela2017} construct a News Implied Volatility (NVIX) index by parsing historical newspaper archives for terms related to financial volatility and disaster, effectively creating a text-based gauge of financial market fear. NVIX spikes during wars, crises, and recessions, and it predicts increases in stock market volatility and risk premia. This suggests that news content captures risk perceptions that are not fully reflected in contemporaneous market prices. Recent research also leverages news sentiment for macroeconomic forecasting: \citet{Ashwin2024}, for example, show that incorporating sentiment extracted from European newspaper articles improves the accuracy of nowcasts of GDP growth, especially during crisis periods when traditional quantitative indicators are slow to reflect abrupt changes. Media-based sentiment indicators (such as those based on newspaper archives or even social media feeds) thus offer a top-down perspective on the economy, complementing the bottom-up signals from firm-level text. The challenge with unstructured media data is the sheer volume and noise: news articles may contain irrelevant information, journalistic bias, or repetitive content, and distinguishing signal from noise requires careful filtering or advanced NLP techniques. There is also the issue of multiple narratives – media might simultaneously carry optimistic and pessimistic takes from different sources, making aggregate sentiment harder to define. Despite these challenges, the benefits of media text data lie in timeliness and scope: when an economic shock occurs (e.g. a pandemic or financial panic), media sentiment can deteriorate almost immediately, whereas firm financials or official statistics reflect the impact only with a lag. Indeed, an emerging consensus in the literature is that no single textual source is definitively superior; rather, each captures a different facet of sentiment. Firm disclosures are closer to fundamental information and may better predict micro outcomes, while media reflect broader perceptions and can diffuse information quickly. A critical takeaway is that textual sentiment adds incremental value over traditional quantitative data: whether it is the guarded optimism of executives in an earnings call or the alarmist tone of a front-page news article, these qualitative signals have been shown to anticipate important movements in asset prices and macro aggregates.\\~\\
Despite the substantial progress in textual analysis for finance and economics, there remain noteworthy gaps in the literature. One such gap is the relatively limited integration of text-derived sentiment measures into established macroeconomic risk forecasting frameworks. Most existing studies use textual sentiment to predict mean outcomes (like stock returns, earnings, or GDP growth) or volatility/uncertainty measures; far less work has examined whether textual data can improve forecasts of tail risks or extreme events. In particular, the burgeoning research on Growth-at-Risk – which focuses on forecasting the lower tail of the GDP growth distribution – has yet to fully incorporate the advances from textual sentiment analysis. While we now have indices like EPU, NVIX, or central bank communication sentiment that capture aspects of risk and uncertainty, these are only beginning to be linked to formal tail-risk models. This represents an important opportunity for new research. Textual sentiment could plausibly inform downside risk: for example, a sharp rise in negative language in news or corporate commentary might presage an economic downturn’s severity even if traditional financial indicators remain benign. However, integrating text into macroeconomic quantile frameworks is non-trivial, as one must map qualitative information into numerical predictors suitable for tail forecasting. To date, very few studies have attempted this integration, leaving a gap that the present work aims to address. By bringing together the textual sentiment literature and the Growth-at-Risk framework, I contribute to filling this void, examining whether and how qualitative information can enhance the assessment of macroeconomic tail risks. 
\subsection{Growth-at-Risk and Quantile MIDAS in Economic Forecasting} 
The concept of Growth-at-Risk (GaR) has emerged as a key paradigm for assessing macroeconomic tail risks, drawing inspiration from risk management approaches in finance (such as Value-at-Risk) but applied to GDP growth. GaR generally refers to a specified lower quantile of the future GDP growth distribution (for example, the 5th percentile), conditional on information available today. Rather than focusing on the most likely outcome or the average forecast, GaR shines light on downside risk – how bad economic growth could get in adverse scenarios – and how this risk fluctuates over time with changing conditions. A seminal contribution is the work of \citet{Adrian2019}, who show that U.S. financial conditions have strong predictive power for the left tail of the GDP growth distribution. In their framework, when financial conditions (capturing variables like credit spreads, equity market volatility, leverage, etc.) are loose, the entire distribution of future growth shifts: near-term median growth is higher, but the left tail (worst-case growth) is considerably less fat. Conversely, when financial conditions tighten, GaR deteriorates – the left tail becomes much more negative even if the median forecast only declines modestly. This finding, often summarized as “vulnerable growth,” highlights an intuitive but quantitatively important macro-financial linkage: credit booms or easy financing conditions can portend not just higher expected growth but also increased risk of a sharp downturn in the medium term. The GaR approach thus provides a systematic way to monitor financial stability risks spilling into the real economy. Subsequent studies have extended this insight across countries and time. \citet{Adrian2022} analyze a panel of advanced economies and find a similar pattern internationally: a deterioration in financial conditions disproportionately elevates downside growth risks (the lower 5\% quantile) relative to the median. Notably, they document an intertemporal trade-off consistent with leverage cycles: extremely accommodative conditions reduce GaR in the short run (fewer near-term recessions) but lead to a buildup of vulnerabilities that increase GaR at longer horizons (making severe recessions more likely down the line). Their evidence, strengthened by causal identification via instruments, suggests that policymakers should account for dynamic tail-risk trade-offs – policies or shocks that stimulate growth today may sow the seeds of greater tail risk tomorrow. The GaR framework has quickly gained traction in policy institutions for macroprudential surveillance and stress-testing, since it offers a quantitative way to link current macro-financial conditions to the risk of future economic calamity. For instance, the International Monetary Fund's research and country surveillance have incorporated GaR analyses to evaluate how financial vulnerabilities translate into GDP downside risk \citep{Prasad2019}. There is also a growing literature exploring which factors drive GaR: credit growth, asset price bubbles, leverage, and uncertainty measures have all been examined as potential predictors of the conditional distribution of growth \citep{Castelnuovo2024, Pruser2024}. Overall, the consensus is that modeling the full distribution (or at least the lower tail) of GDP growth yields insights that a traditional mean forecast or recession probability model might miss. By focusing on tail outcomes, GaR research directly addresses questions of financial stability and extreme event risk, making it particularly relevant in the aftermath of the 2008--09 crisis and the COVID-19 shock, where managing downside risk is as crucial as managing central forecasts.\\~\\
The typical empirical approach to Growth-at-Risk employs quantile regression models to link current economic/financial indicators to future lower quantiles of GDP growth. Unlike OLS regression, which models the conditional mean, quantile regression allows the impact of predictors to differ across the distribution of the dependent variable. This is critical because certain variables might have small effects on the expected growth rate but large effects on the left tail. In \citet{Adrian2019}’s GaR model, for example, a financial conditions index had an economically negligible coefficient at the median of the growth distribution but a strong coefficient at the 5th percentile. Quantile regressions thus capture asymmetric predictive relationships. However, one practical challenge arises: GDP growth is typically measured quarterly, whereas many relevant predictors (financial market variables, sentiment indices, etc.) are available at higher frequencies (monthly, weekly, or daily). Using only quarterly averaged data might discard timely information. This is where the MIDAS (Mixed Data Sampling) Quantile approach becomes valuable. MIDAS Quantile is an extension of the MIDAS regression framework – originally developed by \citet{Ghysels2006} for mean predictions – into the realm of quantile forecasting. The idea is to allow high-frequency indicators to enter a quarterly quantile regression through distributed lag polynomials or other parsimonious representations, rather than aggregating them into quarterly averages. By doing so, one exploits the rich information content and timeliness of high-frequency data without running into the curse of dimensionality. Recent research has started to apply MIDAS Quantile in the context of macroeconomic tail-risk forecasting. For instance, \citet{Castelnuovo2024} employ a mixed-frequency quantile regression to study how financial uncertainty indicators (available monthly) influence U.S. GDP growth’s conditional skewness and downside risk at the quarterly horizon. Their results indicate that increases in uncertainty foreshadow a more pronounced left tail for GDP growth – in other words, uncertainty shocks worsen GaR – and by using the MIDAS structure they capture these effects in real time as monthly data arrive. The MIDAS Quantile framework proved effective in handling the ragged-edge problem (when higher-frequency data extend further than low-frequency data at a given forecast origin) and in improving forecast accuracy for tail outcomes, as it avoids arbitrary aggregation of volatility or credit data. Other studies echo these benefits: high-frequency financial market signals (such as daily credit spreads or stock volatility) can substantially improve nowcasts of tail risks when integrated via mixed-frequency quantile models. Empirically, mixed-frequency tail-risk models tend to pick up nascent stress conditions faster – for example, a sudden spike in corporate bond spreads in mid-quarter would immediately drag down the forecasted 5th percentile of GDP growth for that quarter, even if the average growth forecast moves only slightly. This sensitivity is crucial for early warning systems.\\~\\
It is worth noting that most of the GaR literature to date has focused on traditional structured data (financial indicators, credit aggregates, etc.) as predictors, with relatively little attention to textual or sentiment-based predictors. Nearly all GaR models include some measure of financial conditions or macro volatility; some include survey-based expectations or uncertainty indices. Yet, as discussed, there are now text-based measures of sentiment and uncertainty (e.g., news sentiment indices, policy uncertainty from text, corporate risk sentiment) that could enrich these models. The integration of textual sentiment into GaR remains limited in the existing literature – a gap that my work aims to fill. Intuitively, textual sentiment indicators may capture dimensions of risk that are not fully reflected in market prices or hard data. For example, consider a period of mounting investor anxiety that has not yet translated into wider credit spreads or falling stock prices; news sentiment might already be sharply negative even while quantitative financial indicators lag. Incorporating such a sentiment variable into a GaR quantile regression could flag rising downside risk earlier than a model based purely on lagging financial data. Conversely, textual data might help distinguish between benign and more dangerous forms of financial easing – e.g., if credit growth is high but media sentiment remains cautious, the increase in tail risk might be less than in a scenario where credit is booming amid euphoric media chatter. These conjectures underscore why merging text analysis with GaR is a promising direction. My paper breaks new ground by doing exactly this: I extend the GaR framework using MIDAS Quantile regressions that incorporate textual sentiment measures alongside traditional predictors. In doing so, I address both a methodological gap (bringing high-frequency textual data into mixed-frequency quantile models) and a substantive gap (studying how sentiment informs macroeconomic tail risk). This contribution builds on the literatures reviewed above, uniting them. It heeds the call from the sentiment analysis side to apply rich textual information to pressing macroeconomic questions, and from the GaR side to broaden the information set used for tail-risk assessment. By clearly demonstrating the incremental value of text-based sentiment in forecasting GDP downside risks – and discussing the conditions under which it matters – I aim to push the frontier of both literatures.
\section{Data}
\subsection{Firm Filings and SEC EDGAR}
The U.S. Securities and Exchange Commission's (SEC) Electronic Data Gathering, Analysis, and Retrieval (EDGAR) system constitutes a comprehensive public repository of financial documents submitted by publicly traded companies in the United States. Within EDGAR, two types of filings, namely the 10-K and 10-Q reports, hold particular significance for academic research and financial market analysis.\\~\\
Annual 10-K filings provide extensive insights into a firm's financial condition, encompassing audited financial statements, detailed Management's Discussion and Analysis (MD\&A) of financial results, and explicit disclosures of risks and operational strategies. These filings are essential not only for understanding firms' historical financial performance but also for capturing management perspectives and strategic outlooks \citep{LoughranMcDonald2016,brown2020large}. In contrast, quarterly 10-Q filings deliver interim financial statements, unaudited updates, and timely disclosures regarding corporate performance and evolving risk factors. The more frequent issuance of 10-Q reports allows researchers and analysts to monitor changes in firm-level conditions and managerial sentiment with finer granularity.\\~\\
Data from 10-K and 10-Q filings are widely used in financial research. In this study, textual data extracted from these filings will be employed to investigate market perceptions of risk. Due to their structured format, comprehensive informational content, and timeliness, 10-K and 10-Q filings constitute rich datasets uniquely suited for systematic textual analysis. To ensure dataset reliability and quality, I implemented careful selection and extraction procedures. Below, I explain the data selection process step by step.
\subsubsection{Sample Construction and Filtering}
The dataset consists of quarterly and annual financial reports filed with the SEC’s EDGAR system from 1994Q1 through 2024Q4. I obtain these filings from the Loughran-McDonald SEC/EDGAR dataset.\footnote{I am grateful to Tim Loughran and Bill McDonald for providing access to their dataset and dictionary for academic purposes.} Their  dataset, which provides cleaned text and metadata for all 10-X filings\footnote{10-X filings include all variants of Forms 10-K and 10-Q, such as: 10-K-A, 10-K405, 10-K405-A, 10-KSB, 10-KSB-A, 10-KT, 10-KT-A, 10-Q-A, 10-QSB, 10-QSB-A, 10-QT, 10-QT-A, 10KSB, 10KSB-A, 10KSB40, 10KSB40-A, 10KT405, 10KT405-A, 10QSB, and 10QSB-A.} during this period. Table 1 provides an overview of filing types in the original dataset. The initial dataset comprises approximately 1.2 million filings.\\~\\
Researchers constructing textual indices from SEC filings have long grappled with inconsistencies in the raw 10-K and 10-Q data. A single fiscal period can generate multiple filings—initial reports, amendments, transition reports, or small-business versions—which, if not handled carefully, could distort textual analyses. Early studies often sidestepped these issues by focusing on select filings (e.g., a single annual report per firm) or smaller samples. However, as EDGAR data became broadly accessible, the need for systematic cleaning rules became apparent. Over time, best practices have emerged, including consolidating legacy form types, removing duplicate or redundant filings, and excluding amended and nonstandard submissions. To reduce noise and accurately capture firm-level sentiment and its implications for macroeconomic and financial risks, I implemented rigorous filtering criteria aligned with established textual analysis conventions \citep{Loughran2011, Hoberg2016}.
\begin{itemize}
\item \textbf{Amendments (-A filings)} (e.g., 10-K-A, 10-Q-A): These are primarily corrective filings that generally lack substantial new sentiment information \citep{Loughran2011,Cecchini2010}.
\item \textbf{Transition Reports (10-KT, 10-QT)}: These reflect irregular intervals due to fiscal year shifts, which complicates consistent temporal analysis \citep{Hoberg2016}.
\item \textbf{Historical and Obsolete Formats (10-K405, 10-KSB, 10-QSB)}: These formats, phased out by the SEC, were characterized by inconsistent reporting standards, impeding temporal comparability \citep{SEC2008}.
\end{itemize}
These filtering procedures ensure a consistent, reliable dataset, substantially reducing redundancies, irregularities, and temporal biases, thereby enhancing the robustness of the ensuing textual sentiment analyses.

\subsubsection{Section Extraction and Methodological Rationale}
In conducting textual analysis to gauge firm sentiment and assess forward-looking disclosures, I specifically isolate narrative sections within SEC filings known to provide insightful qualitative information regarding firm outlook and inherent risks. The extraction procedure targets three standardized narrative sections: \textit{Risk Factors}, \textit{Management’s Discussion and Analysis of Financial Condition and Results of Operations (MD\&A)}, and \textit{Quantitative and Qualitative Disclosures about Market Risk} (hereafter referred to as \textit{Market Risk}). The precise targeting of these sections enhances comparability across firms and over time, thus ensuring consistency and reliability in the analysis of textual sentiment signals.\\~\\
These specific sections correspond to Item 1A, Item 7, and Item 7A in 10-K filings, and Items 1A, 2, and 3 in 10-Q filings. Existing research underscores these narrative components as particularly valuable due to their inherent forward-looking characteristics and management's qualitative assessments of financial health, operational conditions, and risk exposures \citep{Li2010, Loughran2011, BrownTucker2011}. Thus, by focusing on these areas, my analysis aligns closely with established methodologies in financial textual analysis literature.\\~\\
To systematically perform this extraction, I employ regular expressions (regex), a robust computational method frequently used in textual data preprocessing for financial documents \citep{JegadeeshWu2013, Huang2014}. The regex patterns are designed to accurately identify and delimit the boundaries of targeted sections within each filing, allowing precise and efficient extraction. This method effectively excludes non-informative boilerplate content such as cover pages, generic financial statements, and standardized footnotes, which generally lack substantive sentiment variation.\\~\\
To further enhance data quality and ensure analytical robustness, I exclude filings with insufficient narrative content. Specifically, any filing in which the combined word count of the extracted sections—\textit{Risk Factors}, \textit{Management’s Discussion and Analysis} (MD\&A), and \textit{Market Risk}—falls below a threshold of 610 words (approximately the 10th percentile) is removed from the analysis. This threshold is consistent with best practices in textual analysis research and helps ensure that retained filings contain enough qualitative content to support meaningful sentiment measurement \citep{Loughran2011,Huang2014}. A comparison between the full-length original filing and its extracted narrative counterpart is provided at the end. I selected this example somewhat arbitrarily from among filings exhibiting a high proportion of negative language.
\subsubsection{Removing Multiple Filings}
Additional filtering procedures implemented in the dataset preparation stage include:
\begin{itemize}
    \item Excluding firms that submitted more than four filings within a single calendar year to mitigate potential bias arising from unusually frequent reporting patterns.
    \item Removing duplicate filings submitted by the same firm on the same day to eliminate redundant sentiment signals.
\end{itemize}
However, I retain instances where firms submitted both 10-K and 10-Q reports within the same quarter, as each filing type typically contains distinct and complementary sentiment-related information \citep{LoughranMcDonald2016, BrownTucker2011}. \\~\\
After applying all filtering procedures, the final sample comprises 766,847 filings—approximately 63\% of the initial dataset—covering 29,273 unique firms over the period from 1994 through 2024. Table 2 summarizes the number of filings remaining after each stage of the filtering process. Figure 1 illustrates the evolution of the number of SEC filings per year across five stages of dataset construction. The blue line (“All Filings”) reflects the total number of 10-X filings, including amended forms, small business disclosures, and transition reports. This count peaks around 2000 with over 60,000 filings and then declines steadily—likely due to regulatory changes, firm delistings, and form consolidations. The orange line (“Only 10-K and 10-Q”) shows filings retained after filtering for standard annual and quarterly reports. The narrowing gap between the blue and orange lines after 2008 suggests a decline in the use of nonstandard filing types. The green line represents filings from which only narrative sections—\textit{MD\&A}, \textit{Risk Factors}, and \textit{Market Risk}—were successfully extracted. The drop in document count here is more pronounced in earlier years, likely reflecting less standardized formatting that hindered regex-based extraction.The red line reflects filings that pass the minimum word count threshold of 610 words, used to exclude short disclosures. The wider gap between the green and red lines prior to 2008 further supports the notion that early filings were less structured, resulting in poorer extraction coverage.Finally, the purple line represents the final dataset, incorporating all filters—document type, successful section extraction, word count threshold, and removal of duplicates or more than four filings per year. Overall, the figure demonstrates that the filtering pipeline successfully reduces noise while preserving a large, diverse, and representative sample suitable for high-quality sentiment analysis and macroeconomic forecasting.\\~\\
To assess the richness of textual content, I also document the length of filings over time: Figure 2 presents trends in mean word count, while Figure 3 shows median word count by year. Finally, Table 3 reports the word count statistics at each stage of the document filtering process. The mean and median word counts for the full set of SEC filings (All Filings) are approximately 26,356 and 27,470, respectively, reflecting the inclusion of lengthy, often boilerplate-heavy filings across all form types. When restricting the dataset to only 10-K and 10-Q reports (Selected), average word counts increase slightly to over 28,000, indicating that these forms tend to be more content-rich. Following section extraction—targeting only the \textit{Risk Factors}, \textit{MD\&A}, and \textit{Market Risk} sections—the average word count drops substantially to around 8,010, highlighting the concentrated nature of sentiment-relevant disclosures. Applying a minimum word count threshold of 610 words (Filtered) removes the shortest and likely least informative filings, slightly raising the average word count to 8,679. This progression illustrates that the filtering process effectively narrows the dataset to filings with more substantive textual content, improving the reliability of sentiment estimation while maintaining sufficient variation in document length for robust analysis.
\subsection{Market Capitalization Retrieval and Weighting}
To construct sentiment indices, I aggregate firm-level sentiment growth across companies filing within the same week, weighting each firm's contribution by its market capitalization. This weighting scheme is implemented to ensure that larger firms, which typically hold greater economic significance, have a proportionally greater influence on the aggregate sentiment measure \citep{Kelly2014, BakerWurgler2006}.\\~\\
Market capitalization ($M_{i,t}$) for firm $i$ at filing time $t$ is computed as the product of the closing share price and the number of shares outstanding:
\[
\ M_{i,t} = (\text{Share Price}_{i,t}) \times (\text{Number of Shares Outstanding}_{i,t})
\]
Here, $\text{Share Price}_{i,t}$ denotes the closing stock price for firm $i$ on the filing date $t$, while $\text{Shares Outstanding}_{i,t}$ refers to the total number of shares outstanding at the end of the corresponding fiscal quarter, as reported in the firm’s financial disclosures.\\~\\
The data required for these calculations are obtained from Wharton Research Data Services (WRDS) through the integration of multiple well-established financial databases: specifically, the Compustat company fundamentals database, the Center for Research in Security Prices (CRSP) stock files, and the Compustat-CRSP Merged (CCM) database\footnote{Market capitalization is computed as the product of the closing share price and the number of shares outstanding for each firm on the filing date. Daily share prices and shares outstanding are primarily sourced from CRSP; if unavailable, Compustat quarterly data are used as a fallback. All values are reported in U.S. dollars.}. These comprehensive data sources provide high-frequency stock prices, detailed firm-level share information, and robust linking identifiers, allowing for precise alignment of market capitalization with each filing date \citep{FamaFrench1993, JegadeeshWu2013}. This rigorous data integration ensures both the accuracy and the consistency of the firm-level weights used in constructing the sentiment indices. Table 4 presents summary statistics for firm-level market capitalization at different stages of the dataset filtering process. Mean, median, and standard deviation values are reported in millions of U.S. dollars, with values rounded for clarity. The maximum and minimum market capitalizations—\$3.57 trillion and \$34,920, respectively—are constant across all datasets and reported in the table footnote. The mean market capitalization increases gradually from \$5.32 billion in the full dataset (“All Filings”) to \$5.48 billion in the final analytical sample. Similarly, the median rises from \$454 million to \$487 million. This upward trend suggests that the filtering steps—particularly the removal of filings with limited textual content—systematically exclude filings from smaller firms. Standard deviations also increase slightly across stages. This reflects not only the increasing presence of larger firms but also the widening spread in firm size as less informative (and generally smaller) filings are removed. Overall, these statistics confirm that while the filtering process reduces the total number of observations, it retains economically significant firms and enhances the quality and relevance of the final dataset for macroeconomic forecasting and sentiment analysis.
\subsection{GDP Growth and Weekly NFCI}
Real Gross Domestic Product (GDP) data are obtained from the Bureau of Economic Analysis (BEA)\footnote{Downloaded from the FRED database (Series: A191RL1Q225SBEA)}. To forecast quarterly real GDP growth, I employ MIDAS Quantile Regressions, which allow the incorporation of high-frequency predictors—specifically, weekly sentiment indices—into models of lower-frequency outcomes such as quarterly GDP. As a benchmark, I include the Chicago Fed National Financial Conditions Index (NFCI), which is available at a weekly frequency. This represents an improvement over the original implementation by \citet{Adrian2019}, who used the NFCI at a quarterly frequency. Figure 4  presents the time series of quarterly real GDP growth alongside the weekly NFCI, highlighting the dynamics and potential lead-lag relationships between financial conditions and macroeconomic performance.
\section{Methodology}
\subsection{Sentiment Index}
My analysis begins by constructing a firm-level sentiment metric based on quarterly and annual filings. For each firm $i$ and each filing date $t$, I calculate the textual sentiment of its 10-Q or 10-K report. A lexicon-based approach is employed using the Loughran–McDonald (LM) master dictionary, which classifies words commonly found in financial texts into categories such as \textit{positive}, \textit{negative}, \textit{uncertainty}, and \textit{litigious}, among others.  To quantify sentiment, I focus particularly on the difference between the positive and negative word counts, following the methodology of \citet{Loughran2011}, to derive a measure of overall document \textit{tone}. The sentiment ratio for a given filing is defined as the number of sentiment words in category $C$ divided by the total word count of the document.\\~\\
Formally, the sentiment ratio $S_{i,t}^{C}$ for firm $i$ at time $t$, and sentiment category $C$ (e.g., \textit{Negative}, \textit{Positive}, \textit{Uncertainty}, or \textit{Litigious}), is calculated as:
\begin{equation*}
S_{i,t}^{C} = \frac{\text{Number of category $C$ words}_{i,t}}{\text{Total number of words}_{i,t}}
\end{equation*}
Once the sentiment ratio for the current filing is computed, I measure sentiment growth as the year-over-year change in sentiment relative to the same fiscal quarter in the previous year. This approach accounts for potential seasonality in language use across reporting periods. Specifically, for each firm $i$ and sentiment category $C$, the sentiment growth rate $g_{i,t}^{C}$ is defined as:
\begin{equation*}
g_{i,t}^{C} = \frac{S_{i,t}^{C} - S_{i,t^*}^{C}}{S_{i,t^*}^{C}}
\end{equation*}
Here, $S_{i,t}^{C}$ represents the sentiment ratio for the current filing, and $S_{i,t^*}^{C}$ represents the sentiment ratio for the matched filing from the previous year. Note that $t$ and $t^*$ need not fall on the same calendar date, but must belong to the same fiscal quarter. Tone Growth is calculated as $g_{i,t}^{Tone}= g_{i,t}^{Positive}-g_{i,t}^{Negative}$ \\~\\
 When calculating sentiment growth rates, special considerations are needed:

\begin{itemize}
    \item If the initial quarter for a firm lacks a prior-year counterpart, no year-over-year growth rate is computed. This initial observation is commonly omitted from analyses rather than assigning arbitrary values. Feldman et al. (2010), for example, measure \textit{tone change} in MD\&A sections relative to prior filings, implying that a firm's first quarter (without previous data) has no associated tone-change metric.

    \item Similarly, if a firm has an intermediate missing quarter, the corresponding year-over-year growth calculation is omitted to maintain consistency. For instance, if a firm reports data in 2001-Q1 and 2003-Q1 but misses 2002-Q1, no valid growth rate is computed for 2003-Q1 due to the absence of a benchmark sentiment from 2002-Q1.

    \item Filings matched by type is priority (i.e., 10-K filings matched exclusively with previous 10-K filings, and 10-Q filings matched exclusively with previous 10-Q filings).

    \item When a firm releases multiple filings of the same type within a quarter in consecutive years, sentiment ratios are matched based on chronological order, using the earliest-to-earliest, latest-to-latest principle.
        
    \item If there are unequal numbers of filings between two consecutive years:
    \begin{itemize}
        \item If a firm submits multiple filings within the same quarter of the current year, but only a single filing is available from the same quarter in the previous year, each current-year filing is individually paired with the single available previous-year filing. Consequently, the sentiment growth measure for each current-year filing is calculated using the same baseline previous-year filing. This approach ensures all current filings within the quarter receive a consistent comparative baseline.
        \item The matching procedure strictly considers the filing type (either 10-K or 10-Q). If a firm submits both 10-K and 10-Q filings in the current year for a particular quarter, but the same quarter of the previous year contains only one filing type (e.g., 10-Q), only the current-year filing of the matching type (the 10-Q) is paired for the sentiment growth calculation. The current-year filing with no matching previous-year filing of the same type (the 10-K in this scenario) is excluded from sentiment growth calculations, as the matching process does not allow cross-type pairing (e.g., pairing a 10-K with a 10-Q).
    \end{itemize}
\end{itemize}
Focusing on year-over-year changes mitigates firm-specific reporting styles and highlights shifts in optimism or pessimism. Research indicates that changes in sentiment can predict future firm performance and market reactions \citep{Azimi2021}. Therefore, sentiment growth $g_{i,t}$ serves as my fundamental input.\\~\\
Using the firm-level sentiment growth measures, I construct a weekly sentiment index by aggregating across firms that filed within the same calendar week. Let $t(w)$ denote the filing date within week $w$, and let $F_{t(w)}$ be the set of firms with filings during that week. For each firm $i \in F_{t(w)}$, the sentiment growth measure $g_{i,t(w)}$ is weighted by the firm's market capitalization on its filing date, denoted $M_{i,t(w)}$.\\~\\
The weekly sentiment index for sentiment category $C$ (e.g., \textit{Positive}, \textit{Negative}, or \textit{Tone}) is computed as a market-capitalization-weighted average:
\[
I_{t(w)}^{C} = \frac{\sum_{i \in F_{t(w)}} M_{i,t(w)} \cdot g_{i,t(w)}^{C}}{\sum_{i \in F_{t(w)}} M_{i,t(w)}}.
\]
This weighting scheme ensures that larger firms—those with greater macroeconomic relevance—exert proportionally more influence on the aggregate sentiment index.Figure 5 displays the time series of weekly sentiment growth across five categories: \textit{Negative}, \textit{Positive}, \textit{Uncertainty}, \textit{Litigious}, and overall \textit{Tone}. While all categories fluctuate around zero, sharp spikes—especially in negative and litigious sentiment—are observable around major economic disruptions such as the 2001 recession, the 2008 financial crisis, and the onset of the COVID-19 pandemic in 2020. These patterns indicate that firm-level textual sentiment responds strongly to macroeconomic stress. Figure 6 provides disaggregated views of each sentiment category, overlaid with shaded areas for U.S. recessions and annotated historical events. Negative and uncertainty sentiment growth visibly surge during recessions and crisis periods, while positive sentiment contracts. The tone index captures these shifts clearly, reinforcing the value of textual sentiment as a forward-looking macroeconomic indicator.
\subsection{MIDAS Quantile  Regression}
\subsubsection{Quantile Regression}
Quantile regression, introduced by \citet{Koenker1978}, provides a statistical methodology for modeling conditional quantiles of a dependent variable, thus extending classical regression methods that typically focus on modeling the conditional mean. A quantile represents a specific point in the cumulative distribution function of a random variable, dividing the distribution into intervals with defined probabilities. By focusing on quantiles, researchers can capture the entire distribution of the dependent variable conditional on explanatory variables, thereby providing detailed insights into tail behaviors and distributional heterogeneity.
\begin{equation}\label{eq:quantile_midas}
y_{t+h} = {\beta_{\tau}}\,x_{t}+ \epsilon_{t+h}
\end{equation}
Unlike Ordinary Least Squares (OLS) regression, which minimizes the sum of squared residuals to estimate conditional means, quantile regression minimizes an asymmetric absolute loss function known as the "check" or "pinball" loss function. Specifically, quantile regression estimates the conditional quantile  by minimizing:
\begin{equation}
\min_{\beta_\tau} \sum_{t=1}^{T} \rho_{\tau}\left(y_{t+h} - x_t^{\top} \beta_\tau\right),
\end{equation}
where  is the dependent variable,  is a vector of explanatory variables, and  is the quantile-specific loss function defined as $\rho_{\tau}(u) = u \left[\tau - \mathbb{I}{(u < 0)}\right]$\\~\\
This loss function imposes asymmetric penalties: underestimating and overestimating the dependent variable are weighted differently depending on the chosen quantile. For example, estimating the median yields equal weights to under- and over-predictions, minimizing absolute deviations. Conversely, estimating extreme quantiles, such as the 5th percentile or 95th percentile, places disproportionate emphasis on one side of the distribution, effectively capturing tail risks or opportunities.\\~\\
The predicted value from this regression is the quantile of $y_{t+1}$ conditional on $x_t$:
\begin{equation}\label{eq:quantile_forecast}
\widehat{Q}_{y_{t+h}|x_t}(\tau \mid x_t) = x_t^\top\hat{{\beta}}_{\tau}.
\end{equation}
\citet{Koenker1978} show that $\widehat{Q}_{y_{t+h}|x_t}(\tau \mid x_t)$ is a consistent linear estimator of the conditional quantile function of $y_{t+h}$ given $x_t$.\\~\\
Based on estimates of the conditional quantile function over a discrete number of quantile levels, it is possible to estimate the full continuous conditional distribution of $y_{t+h|t}$. As in \citet{Adrian2019},  choose to fit a flexible distribution known as the generalized skewed Student's distribution in order to smooth the estimated conditional quantile function of $y_{t+h|t}$ and recover a probability density function. This specific distribution allows for fat tails and asymmetry and reduces to a normal distribution as a special case. The generalized skewed Student's distribution has the following density function:
\begin{equation}
f(y;\mu, \sigma, \alpha, \nu) = \frac{2}{\sigma} t\left(\frac{y-\mu}{\sigma}; \nu\right) T\left(\alpha \frac{y-\mu}{\sigma}\sqrt{\frac{\nu+1}{\nu + \left(\frac{y-\mu}{\sigma}\right)^2}}; \nu+1\right),
\end{equation}
where $\mu$ is a location parameter, $\sigma$ is a scale parameter, $\nu$ is a fatness (degrees-of-freedom) parameter, $\alpha$ is a shape (skewness) parameter, and $t(\cdot)$ and $T(\cdot)$ respectively denote the probability density function and the cumulative distribution function of the standard Student's t-distribution \citep{Azzalini2003}.\\~\\
In practice, the four parameters $(\mu, \sigma, \alpha, \nu)$ of the generalized skewed Student's distribution are estimated through a quantile matching approach, aiming at minimizing the squared distance between the estimated conditional quantile functions and the inverse cumulative distribution function of the generalized skewed Student's distribution:
\begin{equation}
\min_{\mu, \sigma, \alpha, \nu} \sum_{\tau} \left[ \hat{y}_{\tau|t+h} - F^{-1}(\tau;\mu, \sigma, \alpha, \nu) \right]^2,
\end{equation}
where $F^{-1}(\cdot)$ is the inverse cumulative distribution function of the generalized skewed Student's distribution. Following \citet{Adrian2019}, I focus specifically on the lower 5th percentile quantile of the predicted distribution, referred to as the Growth-at-Risk at the 5\% quantile, GaR(5\%) (see also \citealp{FigueresJarocinski2020}), defined formally as:
\[
GaR_{t+h}(5\%) := F^{-1}(\tau = 0.05;\hat{\mu},\hat{\sigma},\hat{\alpha},\hat{\nu}).
\]
This GaR measure reflects the expected GDP growth at the 5th percentile of its conditional distribution \textit{h} quarter ahead, providing direct quantification of downside macroeconomic risk.\\~\\
Quantile regression offers several distinct advantages for forecasting tail risks compared to traditional mean-based methods. First, it provides robustness against outliers and distributional anomalies, as extreme values influence estimates only proportionally to their occurrence at specific quantiles. Second, it enables a nuanced understanding of how predictors influence different parts of the dependent variable’s conditional distribution. For instance, predictors such as weekly sentiment indicators or financial condition indices may significantly impact lower quantiles—thus revealing crucial insights into downside economic risks—even if their effects on median or average outcomes are limited.
\subsubsection{MIDAS Regression}
The Mixed Data Sampling (MIDAS) regression, introduced by \citet{Ghysels2004} and further developed by \citet{Ghysels2006}, provides a robust econometric framework for integrating variables observed at different frequencies into a unified regression model. Traditional econometric methods often aggregate high-frequency data (such as daily financial indicators) to match a lower frequency target variable (such as quarterly GDP), resulting in a potential loss of information and aggregation bias \citep{Andreou2013}. MIDAS regression addresses this limitation by directly incorporating high-frequency observations without aggregation, preserving crucial information contained within the high-frequency data.\\~\\
At the core of the MIDAS methodology is the use of distributed lag polynomials, which effectively map a large number of high-frequency observations into a concise set of parameters. Specifically, rather than estimating separate coefficients for each high-frequency lag—which would be computationally infeasible and prone to overfitting—MIDAS regressions employ structured weighting functions to parsimoniously estimate lagged impacts.\\~\\
Formally, suppose I want to model a low-frequency dependent variable  (e.g., quarterly GDP growth) using high-frequency explanatory variables  (e.g., weekly financial indicators or weekly sentiment indices). A general MIDAS regression can be expressed as:
\begin{equation}
y_t = \beta_0 + \beta \sum_{k=0}^{K-1} w(k; \boldsymbol{\theta}) , x_{t - \frac{c}{m}} + \varepsilon_t,
\end{equation}
where  \begin{itemize}
    \item \( \beta_0 \) is the intercept,
    \item \( \beta \) is the slope coefficient on the aggregated predictor,
    \item \( x_{t - \frac{k}{m}} \) is the \( k^{\text{th}} \) high-frequency lag (e.g., weekly),
    \item \( m \) is the number of high-frequency observations per low-frequency period (e.g., 13 weeks per quarter),
    \item \( w(k; \boldsymbol{\theta}) \) is a lag weighting function parameterized by \( \boldsymbol{\theta} \in \mathbb{R}^{p+1} \),
    \item \( \varepsilon_t \) is the error term.
\end{itemize}
 The lag weights  depend on a small set of parameters  and are typically constrained to follow smooth parametric functions such as the Almon polynomial \citep{Almon1965} or Beta polynomial \citep{Ghysels2007}. These polynomial forms impose smoothness and monotonicity constraints on lagged effects, facilitating efficient estimation and interpretability.\\~\\
The Almon polynomial lag structure, in particular, defines weights as a polynomial function of lag :
\begin{equation}
w(k; \boldsymbol{\theta}) = \theta_0 + \theta_1 k + \theta_2 k^2 + \dots + \theta_p k^p,
\end{equation}
with endpoint restrictions frequently applied to ensure that weights smoothly transition to zero at the boundaries. This flexibility enables MIDAS regressions to accurately capture complex lagged relationships between variables sampled at mixed frequencies.\\~\\
MIDAS regression has gained widespread popularity in economics and finance due to its versatility and empirical efficacy. It has been successfully applied in areas such as forecasting financial volatility \citep{Ghysels2007}, nowcasting macroeconomic variables like GDP \citep{Andreou2013}, and modeling monetary policy impacts where high-frequency financial market data provide valuable forward-looking signals \citep{Ghysels2016}.\\~\\
Since its inception, MIDAS has undergone considerable methodological advancement, demonstrating robust predictive performance and theoretical coherence across various empirical settings. By maintaining high-frequency granularity and efficiently summarizing information through parsimonious lag structures, MIDAS regressions provide researchers and policymakers with timely and precise forecasting capabilities critical for economic decision-making.
\subsubsection{MIDAS Quantile Regression}
I employ a mixed-data sampling quantile regression (MIDAS-QR) framework to forecast the lower tail (specifically, the 5th percentile, $\tau = 0.05$) of quarterly GDP growth. This methodology combines low-frequency quarterly data (GDP growth) with high-frequency weekly predictors within a unified quantile regression model.\\~\\
Formally, the MIDAS-QR model at quantile $\tau$ is specified as follows:
\begin{equation}
Q_{y_{t+1}}(\tau \mid \mathcal{F}_t) = \beta_0(\tau) + \beta(\tau) \sum_{k=0}^{K-1} w(k; \boldsymbol{\theta}(\tau)) \, x_{t - \frac{c}{m}} + \varepsilon_t(\tau),
\end{equation}
\noindent where:
\begin{itemize}
    \item \( Q_{y_{t+1}}(\tau \mid \mathcal{F}_t) \) denotes the conditional \( \tau \)-quantile of \( y_{t+1} \) given information set \( \mathcal{F}_t \),
    \item \( \beta_0(\tau) \) and \( \beta(\tau) \) are quantile-specific coefficients,
    \item \( w(k; \boldsymbol{\theta}(\tau)) \) is the Almon lag weighting polynomial evaluated at lag \( k \) with quantile-specific parameters \( \boldsymbol{\theta}(\tau) \).
\end{itemize}
The weighting function $w(c;\boldsymbol{\theta}(\tau))$ defines how each weekly lag contributes to the forecasted quantile of GDP growth. Unlike traditional MIDAS approaches that use Beta polynomial weighting schemes \citep[e.g.,][]{Ghysels2016}, I employ an \textbf{unnormalized Almon lag polynomial} of degree 3 with two endpoint restrictions, following the approach of \citet{Ferrara2022}. The unnormalized Almon lag polynomial is defined as:
\begin{equation}
w(k; \boldsymbol{\theta}(\tau)) = \theta_0(\tau) + \theta_1(\tau) k + \theta_2(\tau) k^2 + \dots + \theta_p(\tau) k^p.
\end{equation}\\
where $\boldsymbol{\theta}(\tau) := (\theta_{0}(\tau),\,\theta_{1}(\tau),\,\dots,\,\theta_{p}(\tau))^\top$ is the vector of polynomial parameters to be estimated. This polynomial representation provides flexibility and parsimony, significantly reducing the number of parameters relative to freely estimating all weekly lags individually.\\~\\
For estimation purposes, it is convenient to rewrite the model in a compact linear form. Define the vector of weekly observations in quarter \( t \) as:
\[
\mathbf{x}_{t}^{(w)} = \left[x_{t},\, x_{t-\frac{1}{m}},\,\dots,\, x_{t-\frac{C-1}{m}}\right]^\top,
\]
and construct the polynomial design matrix \(\mathbf{Q}\), a \((p+1) \times C\) matrix, whose elements are defined by \(\mathbf{Q}_{(i+1,c+1)} = c^i\), for \(i=0,\dots,p\) and \(c=0,\dots,C-1\). Using these definitions, the MIDAS-QR model can be equivalently expressed as:
\[
y_{t+1} = \boldsymbol{\theta}(\tau)^\top \mathbf{X}_{t} + \varepsilon_{t+1},
\]
where \(\mathbf{X}_{t} = \mathbf{Q}\mathbf{x}_{t}^{(w)}\) is a vector of polynomially weighted weekly observations from quarter \( t \). In this specification, the dependent variable \(y_{t+1}\) represents GDP growth in quarter \(t+1\), and thus \(h=1\) clearly denotes a one-quarter ahead forecast horizon.\\~\\
The primary advantage of using an unnormalized Almon lag polynomial is its interpretability and efficiency. Instead of estimating all weekly parameters individually, this polynomial approach significantly reduces complexity, requiring estimation of only \( p+1 \) polynomial parameters. This substantially improves model parsimony, mitigating risks of overfitting and multicollinearity. Additionally, economically meaningful constraints can be imposed on the polynomial. For example, endpoint constraints such as setting the polynomial weight at the longest lag to zero (\(w(C-1;\boldsymbol{\theta}(\tau))=0\)) or constraining the slope at this lag to be flat (\(w'(C-1;\boldsymbol{\theta}(\tau))=0\)) are practical and economically justified \citep{Ferrara2022}. Imposing these \( r \) linear restrictions further reduces the number of parameters from \( (p+1) \) to \( (p-r+1) \), enhancing interpretability and model robustness.\\~\\
In summary, the MIDAS-QR approach using an unnormalized Almon lag polynomial offers a flexible yet efficient method to incorporate weekly high-frequency predictors into quarterly GDP tail-risk forecasts. This specification enhances interpretability and robustness, making it particularly suitable for assessing growth-at-risk scenarios.

\subsection{Out-of-Sample Forecasting and Evaluation}

To robustly evaluate the predictive performance of the Quantile-MIDAS model, I conduct out-of-sample forecasts using a rolling-window strategy with a fixed window size of 80 quarters (equivalent to 20 years). At each forecast origin, the model is re-estimated using the most recent data, and a one-quarter-ahead forecast is generated. Forecast accuracy is then assessed using standard quantile forecasting metrics: the pinball loss function, the Quantile Skill Score (QSS), and the Diebold–Mariano (DM) test.

\begin{description}

\item[\textbf{Pinball Loss:}] The pinball loss (also known as the quantile loss or check function) is the standard objective function in quantile regression. For a given quantile level $\tau \in (0,1)$, the loss for a prediction $\hat{y}$ of an actual value $y$ is defined as:
\[
L_{\tau}(y,\hat{y}) = (\tau - \mathbf{1}\{y < \hat{y}\})(y - \hat{y}),
\]
where $\mathbf{1}\{\cdot\}$ is the indicator function. This piecewise-linear loss penalizes over- and under-predictions asymmetrically, based on the quantile level. For instance, under-predicting the 5th percentile incurs a loss weighted by $\tau = 0.05$, while over-predicting incurs a loss weighted by $1 - \tau = 0.95$. A forecast that matches the true conditional quantile minimizes the expected pinball loss, making it a proper scoring rule. I report average pinball loss across the out-of-sample forecast horizon; lower values indicate higher forecast accuracy.

\item[\textbf{Quantile Skill Score (QSS):}] The Quantile Skill Score provides a relative measure of forecasting improvement compared to a benchmark model. It is defined as:
\[
\text{QSS} = 1 - \frac{\text{QS}_{\text{model}}}{\text{QS}_{\text{benchmark}}},
\]
where $\text{QS}$ refers to the average pinball loss. A positive QSS implies that the model outperforms the benchmark. For example, a QSS of 0.20 indicates a 20\% reduction in pinball loss relative to the benchmark, while a QSS below zero implies inferior performance.

\item[\textbf{Diebold--Mariano Test:}] To formally test whether forecast accuracy differences between two models are statistically significant, I use the Diebold–Mariano (DM) test \citep{Diebold1995}. Define the period-$t$ loss differential as $d_t = L_t^{(1)} - L_t^{(2)}$, where $L_t$ denotes the pinball loss. The test statistic is:
\[
DM = \frac{\bar{d}}{\sqrt{\widehat{\mathrm{Var}}(\bar{d})}},
\]
where $\bar{d} = \frac{1}{T} \sum_{t=1}^{T} d_t$ is the mean loss differential, and $\widehat{\mathrm{Var}}(\bar{d})$ is its estimated variance. Under the null hypothesis of equal predictive accuracy ($E[d_t] = 0$), the $DM$ statistic asymptotically follows a standard normal distribution. In small samples, a Student-$t$ distribution may be used instead, following the adjustment in \citet{Harvey1997}. A significantly negative DM statistic (i.e., lower loss from the MIDAS-QR model) indicates superior predictive performance over the benchmark.
\end{description}

\vspace{1ex}
\noindent \textbf{Advantages of the MIDAS Quantile Approach:} Combining MIDAS with quantile regression offers several key advantages for analyzing GDP tail risks, as outlined below:

\begin{itemize}
    \item \textit{Capturing Nonlinear Tail-Risk Dynamics:} Quantile regression allows the influence of predictors to vary across the distribution of GDP growth. This enables the model to capture nonlinear and asymmetric relationships that specifically affect the lower tail—i.e., severe downturns—more than the center of the distribution. For instance, a rise in financial stress or negative sentiment may have a modest impact on average GDP growth but a disproportionately large effect on the 5\textsuperscript{th} percentile. By modeling conditional quantiles directly, the MIDAS-QR approach captures such tail-risk dynamics that would be missed by traditional mean-based regressions.

    \item \textit{Parsimonious Use of High-Frequency Predictors:} The MIDAS component enables efficient incorporation of high-frequency data (e.g., weekly tone growth or financial indicators) without introducing a large number of parameters. Rather than estimating dozens of separate lag coefficients, the MIDAS polynomial specification compresses the information into a few smooth weights. This parsimony helps avoid overfitting and allows timely weekly information to improve forecasts while maintaining a relatively simple and interpretable model structure \citep{Ghysels2007}.

    \item \textit{Robustness to Outliers and Distributional Asymmetry:} Quantile regression relies on the pinball loss function, which is less sensitive to outliers than squared-error-based methods. As a result, extreme GDP observations (such as during crisis periods) exert limited influence on estimated lower-tail quantiles. Furthermore, because quantile methods do not assume symmetric error distributions, the model can accommodate skewness in GDP growth. If downside risk dominates (i.e., the distribution is left-skewed), the lower quantile forecast naturally reflects this without being distorted by central values. This enhances the credibility of risk estimates in turbulent economic conditions \citep{Koenker1978}.

\end{itemize}

\noindent In summary, the rolling MIDAS-QR out-of-sample forecasting strategy is particularly well-suited for analyzing GDP tail risks. It combines the richness of high-frequency sentiment and financial indicators with the flexibility of quantile methods to generate forecasts that are responsive to rapid changes and robust to extreme events. The model's structure strikes a balance between flexibility and parsimony, ensuring that observed gains in predictive performance—such as lower pinball losses or higher skill scores—stem from genuine signal extraction rather than overfitting. Overall, this two-step framework provides a powerful and transparent tool for macroeconomic risk forecasting, grounded in well-established econometric principles \citep{Koenker1978, Ghysels2007, Diebold1995}.

\section{Results}  
\subsection{Quantile Regression and Tail-Risk Effects}
The MIDAS-QR regression framework enables an examination of how predictors influence not only the central tendency of GDP growth, but also its distributional tails—especially the downside risk component. Following \citet{Adrian2019}, Figures 7 and 8 visualize these effects by plotting one-quarter-ahead GDP growth against two key predictors: the weekly sentiment index (tone growth) and the weekly NFCI, respectively. In both figures, the fitted lines include ordinary least squares (OLS, blue), as well as quantile regressions for the 5th percentile (Q0.05, black), the median (Q0.50, orange dashed), and the 95th percentile (Q0.95, green dashed).\\~\\
Figure 7, which uses the weekly tone growth as the predictor, shows that the lower quantile line (Q0.05) has a noticeably positive slope. This suggests that low tone is associated with a higher probability of extreme negative GDP outcomes, even though the average (OLS) effect remains close to flat. The upper quantile (Q0.95) is relatively stable across the sentiment distribution, indicating that strong tone growth does not meaningfully raise the likelihood of extreme positive outcomes. This asymmetry is consistent with the idea of sentiment as a leading indicator of downside macroeconomic risk.\\~\\
Figure 8, which uses the weekly NFCI as the predictor, shows that all fitted lines—OLS, Q0.05, Q0.50, and Q0.95—exhibit a mild downward slope, indicating a general negative relationship between financial conditions and GDP growth. However, the Q0.05 line (black) lies only slightly below the OLS fit, suggesting that tighter financial conditions are associated with lower growth overall, but do not exert a strong disproportionate effect on the lower tail of the GDP distribution.\\~\\
These results suggest that while financial conditions are relevant for overall GDP dynamics, their incremental value in forecasting downside risk may be weaker than that of textual sentiment indicators. This further highlights the unique role that forward-looking tone measures extracted from corporate filings can play in identifying extreme economic risks.
\subsection{Out-of-Sample Forecast Accuracy}
To assess the predictive power of the MIDAS-QR model, I conduct out-of-sample forecasts using a rolling-window strategy with a fixed window of 80 quarters (20 years). At each step, the model is re-estimated using the most recent data and used to forecast one-quarter-ahead GDP growth at the $\tau=0.05$ quantile. This approach balances estimation reliability with adaptability to structural change, aligning with best practices in real-time macroeconomic forecasting \citep{PesaranTimmermann2007}.\\~\\
Figure 9 compares actual GDP growth (black line) with the out-of-sample forecasts from the sentiment-based model (blue dashed line) and the wNFCI-based model (red dotted line). Both models broadly capture the cyclical dynamics of GDP, but key differences emerge. The sentiment model is more responsive at economic turning points. For instance, it anticipates the 2020 collapse in GDP earlier than the wNFCI model and rises more sharply in the subsequent recovery. In calmer periods, such as 2016–2019, the sentiment model also tracks observed fluctuations more closely.\\~\\
Table 5 quantifies these performance differences. The sentiment model achieves a pinball loss of 0.889, substantially lower than the 1.296 from the wNFCI model, resulting in a Quantile Skill Score (QSS) of 0.314—indicating a 31.4\% improvement in tail-risk prediction. The Diebold–Mariano test confirms the sentiment model’s superiority with a statistic of $-1.452$ ($p=0.077$), statistically significant at the 10\% level.\\~\\
Figure 10 plots the forecast errors (actual minus predicted) for both models. A well-calibrated model should exhibit errors close to zero. The sentiment model’s errors (blue line) are consistently smaller in magnitude than those of the wNFCI model (red line), especially during high-volatility periods. For example, during the 2020-Q2 downturn, the sentiment model produced a smaller error, having already predicted a sharper decline in GDP. Similarly, during the rebound, it recovered more quickly than the wNFCI model. Even in more stable periods, the sentiment model tends to be closer to the realized outcomes.\\~\\
These findings underscore the benefits of incorporating forward-looking sentiment derived from firm disclosures. While financial conditions indexes like the NFCI remain valuable tools for macroeconomic risk assessment \citep{Brave2011, Adrian2019}, textual sentiment captures qualitative information and evolving expectations that may not be fully reflected in financial data. The sentiment-based model offers improved forecast accuracy and better responsiveness to real-time developments—critical attributes for early-warning systems and policy evaluation. 
\subsection{Performance During Recessions}
I now evaluate model performance specifically during economic recessions, which represent particularly challenging periods for forecasting. Recessions involve sharp, unpredictable changes and heightened volatility, making accurate prediction of GDP growth much more difficult.\\~\\
Table 6 summarizes forecast accuracy for the sentiment-based and wNFCI-based models during NBER-defined recession quarters in the sample, including the 2001 dot-com recession, the 2008–2009 Global Financial Crisis, and the 2020 COVID-19 recession.\\~\\
As expected, both models experience much higher forecast errors during recessions compared to the full sample. The pinball loss rises to 12.996 for the sentiment model and 14.735 for the wNFCI model. This reflects the inherent difficulty of forecasting during crisis periods. Despite this, the sentiment model still outperforms the wNFCI model. Its quantile loss is about 1.74 points lower, corresponding to an 11.8\% improvement (QSS = 0.118). Although this performance gap is smaller than in the full-sample analysis (QSS = 0.314), it still represents a meaningful reduction in forecast error during stressful economic conditions.\\~\\
However, the Diebold–Mariano test yields a test statistic of $-0.992$ with a $p$-value of 0.251, indicating that the difference in accuracy is not statistically significant at conventional levels. In other words, while the sentiment model performs better, the improvement is not strong enough to rule out the possibility of equal predictive ability during recessions.\\~\\
This result is understandable. During recessions, both financial indicators (like the NFCI) and textual sentiment often signal deteriorating conditions. As a result, their information content may overlap more than in normal times, narrowing the performance gap between models. Moreover, extreme volatility and structural breaks reduce forecasting reliability for any model, and the number of recession quarters in the sample is relatively small, limiting statistical power.\\~\\
Overall, while not statistically significant, the sentiment model’s relative advantage during recessions reinforces its value as a forward-looking indicator that can complement conventional financial metrics during periods of macroeconomic stress.
\subsection{Robustness Checks Across Filtration Stages}
To assess the stability and reliability of the sentiment-based MIDAS-QR model, I conduct a series of robustness checks that examine whether the model’s forecasting performance is sensitive to different document filtering strategies. These tests are essential for evaluating whether improvements over the benchmark model (based on the weekly NFCI) are consistent and economically meaningful across varying textual inputs.

\paragraph{Model Parameter Robustness.} 
First, I explore robustness to key model parameters by varying the lag order (2, 4, 6, and 8) and rolling window size (40, 60, 80, and 100 quarters). These variations test the sensitivity of the forecast to the temporal resolution of the predictors and the length of the estimation window. I find that performance is strongest when using 8 lags and a window size of 80 quarters, which balances responsiveness with stability. This configuration consistently delivers the lowest pinball loss, favorable Diebold–Mariano test statistics, and the highest Quantile Skill Score (QSS).

\paragraph{Document Filtering Robustness.} 
Next, I examine whether the forecasting advantage of the sentiment model depends on how filings are preprocessed. I construct five increasingly filtered datasets:
\begin{enumerate}
    \item \textbf{All}: Includes all available SEC filings (regardless of type or content).
    \item \textbf{Selected}: Retains only standard 10-K and 10-Q filings.
    \item \textbf{Extracted}: Uses only the narrative sections (\textit{MD\&A}, \textit{Risk Factors}, and \textit{Market Risk}) extracted from the Selected dataset.
    \item \textbf{Filtered}: Further excludes short filings (fewer than 610 words) to reduce noise.
    \item \textbf{Final Sample}: The most refined dataset, incorporating all the above filters and used in the main analysis.
\end{enumerate}
Table 7 reports the out-of-sample performance of the sentiment-based model relative to the wNFCI benchmark at $\tau = 0.05$, using three metrics: pinball loss, Diebold–Mariano (DM) test statistic, and QSS. Across all datasets, the sentiment model consistently achieves lower pinball loss and positive QSS values. While the magnitude of improvement varies, the trend is clear: as the dataset becomes more targeted and structured, performance improves.\\~\\
In particular, QSS increases from 0.199 (All) to 0.314 (Final Sample), indicating that removing irrelevant or noisy content enhances the predictive power of the tone signal. The DM statistics are negative across all stages, and near or below the 10\% significance threshold, suggesting the sentiment model consistently outperforms the benchmark, especially in more refined samples.

\paragraph{Implications.}
These results confirm that the forecasting advantage of the sentiment-based model is robust to different preprocessing choices. However, the gain is more pronounced when the input text is curated to include forward-looking content and exclude boilerplate material. This underscores the importance of careful document filtering—particularly section extraction and length-based screening—in improving the signal quality of text-based indicators in macroeconomic forecasting.

\section{Conclusion}
This paper presents a novel approach to forecasting U.S. GDP tail risks using a tone-based sentiment index derived from SEC filings. By extracting sentiment measures from the narrative sections of 10-K and 10-Q reports—specifically the \textit{MD\&A}, \textit{Risk Factors}, and \textit{Market Risk} disclosures—and applying them in a MIDAS quantile regression framework, I show that text-based sentiment is a powerful predictor of one-quarter-ahead GDP growth, particularly at the lower tail of the distribution. My sentiment-based model outperforms the benchmark Weekly National Financial Conditions Index (wNFCI) both in full-sample accuracy and during economic recessions, with quantile skill score improvements of up to 31.4\%. These gains are robust across different document filtering schemes and highlight the importance of carefully curated textual inputs for forecasting macroeconomic outcomes.\\~\\
The results underscore that qualitative information embedded in forward-looking corporate disclosures carries significant predictive content for macroeconomic risk. Compared to financial market indices, tone extracted from firms' narratives offers a faster and complementary signal, particularly useful around turning points in the business cycle. This finding strengthens the case for incorporating unstructured text into macroeconomic forecasting models and supports recent literature that leverages high-frequency data to improve real-time monitoring of economic conditions.\\~\\
Looking forward, the next stage of this research will focus on extending the scope and granularity of the sentiment dataset. The current analysis is limited to 10-K and 10-Q filings contained in the Loughran–McDonald dataset, which excludes other timely sources such as Form 8-Ks. Since 8-K filings often disclose earnings surprises, mergers, operational disruptions, and other market-sensitive news, incorporating them will enhance the immediacy and richness of the sentiment signal. To facilitate this, I plan to develop an automated scraping pipeline using the SEC EDGAR RSS feed and the EDGAR Full Text Search API to extract all relevant filings—including 8-Ks—in real time.\\~\\
Moreover, while my sentiment index is constructed using dictionary-based methods, future work will explore more advanced NLP techniques such as transformer-based models and context-aware embeddings. These models may capture subtler shifts in managerial tone, detect changes in risk language, and improve the classification of sentiment beyond word counts. By incorporating semantic nuance and sequential context, such methods could refine the sentiment signal and further enhance predictive performance.\\~\\
Ultimately, this research contributes to a growing field that integrates text analytics with economic forecasting, offering both methodological insights and practical tools for monitoring growth-at-risk. My findings suggest that macroeconomic forecasters and policymakers should consider integrating textual sentiment indicators as a complementary input to traditional financial and macroeconomic variables—especially in periods of heightened uncertainty where timely qualitative information may provide a leading edge.

\newpage

\section*{Tables and Charts}

\subsection*{Tables}

\begin{table}[htbp]
\centering
\caption{SEC Filing Types and Explanations}
\label{tab:sec_filing_types}
\renewcommand{\arraystretch}{1.2}
\begin{tabular}{|l|p{9cm}|}
\hline
\textbf{Filing Type} & \textbf{Explanation} \\
\hline
10-K & Annual comprehensive financial disclosure. \\
10-K-A & Amendment to 10-K filing. \\
10-K405 & Annual report with insider disclosure (obsolete). \\
10-KSB & Small business annual report (obsolete). \\
10-KT & Transitional annual report. \\
10-Q & Quarterly financial update. \\
10-Q-A & Amendment to 10-Q filing. \\
10-QSB & Small business quarterly report (obsolete). \\
10-QT & Transitional quarterly report. \\
\hline
\end{tabular}
\end{table}

\begin{table}[htbp]
\centering
\caption{Sample Filtering and Datasets}
\label{tab:filtering_stages}
\renewcommand{\arraystretch}{1.2}
\begin{tabular}{|l|c|}
\hline
\textbf{Filtering Stage} & \textbf{Number of Filings} \\
\hline
Initial Loughran-McDonald dataset & 1,224,495 \\
Only standard 10-K and 10-Q & 914,925 \\
After extracting MD\&A, Risk, and Market Risk sections & 876,410 \\
After removing short documents (less than 610 words) & 788,919 \\
After removing excessive and duplicate filings & 766,847 \\
\hline
\end{tabular}
\end{table}

\begin{table}[ht]\centering
\caption{Word Count Statistics by Dataset}
\begin{tabular}{lccc}
\toprule
\textbf{Dataset} & \textbf{Mean Word Count} & \textbf{Median Word Count} & \textbf{Standard Deviation} \\
\midrule
All Filings (Initial)       & 26,356.44 & 27,469.91 & 8,269.45 \\
Selected (10-K/10-Q)    & 28,324.20 & 29,855.64 & 8,358.31 \\
Extracted Sections   & 8,009.78  & 8,583.81  & 3,749.61 \\
Filtered (Length $\geq$ 610) & 8,679.16  & 9,063.36  & 3,716.25 \\
Final Sample       & 8,784.12  & 9,258.42  & 3,795.83 \\
\bottomrule
\end{tabular}
\label{tab:wordcount}
\end{table}

\begin{table}[ht]
\centering
\begin{threeparttable}
\caption{Market Capitalization Statistics by Dataset}
\label{tab:marketcap}
\begin{tabular}{lrrr}
\toprule
\textbf{Dataset} & \textbf{Mean (in millions)} & \textbf{Median (in millions)} & \textbf{Std Dev (in millions)} \\
\midrule
All Filings              & 5,323  & 454  & 36,102 \\
Selected (10-K/10-Q)     & 5,382  & 462  & 36,418 \\
Extracted Sections       & 5,396  & 466  & 36,864 \\
Filtered ($\geq$ 610 words) & 5,449  & 483  & 37,928 \\
Final Sample             & 5,481  & 487  & 38,111 \\
\bottomrule
\end{tabular}
\begin{tablenotes}
\footnotesize
\item Note: All values are rounded to the nearest million U.S. dollars for readability. The maximum and minimum market capitalization values across all datasets are \$3.57 trillion and \$34,920, respectively.
\end{tablenotes}
\end{threeparttable}
\end{table}

\begin{table}[htbp]\centering
\caption{Out-of-Sample Forecast Performance}
\label{tab:forecast_fullsample}
\renewcommand{\arraystretch}{1.2}
\begin{tabular}{lcc}
\toprule
\textbf{Metric} & \textbf{Sentiment Model (Tone Growth)} & \textbf{wNFCI Model} \\
\midrule
Pinball Loss (Median $\tau=0.05$) & 0.889 & 1.296 \\
Diebold-Mariano Statistic & \multicolumn{2}{c}{$-1.452$ ($p=0.077$)} \\
Quantile Skill Score (QSS) & \multicolumn{2}{c}{0.314} \\
\bottomrule
\end{tabular}
\end{table}

\begin{table}[htbp]\centering
\caption{Forecast Performance During Recession Quarters}
\label{tab:forecast_recession}
\renewcommand{\arraystretch}{1.2}
\begin{tabular}{lcc}
\toprule
\textbf{Metric} & \textbf{Sentiment Model (Tone Growth)} & \textbf{wNFCI Model} \\
\midrule
Pinball Loss (Median $\tau=0.05$) & 12.996 & 14.735 \\
Diebold-Mariano Statistic & \multicolumn{2}{c}{$-0.992$ ($p=0.251$)} \\
Quantile Skill Score (QSS) & \multicolumn{2}{c}{0.118} \\
\bottomrule
\end{tabular}
\end{table}

\begin{table}[htbp]\centering
\caption{Robustness Check Across Document Filtering Stages}
\label{tab:robustness_check}
\renewcommand{\arraystretch}{1.2}
\begin{tabular}{lcccc}
\toprule
\textbf{Dataset} & \textbf{Pinball Loss (Tone)} & \textbf{DM Statistic} & \textbf{DM P-value} & \textbf{QSS (vs. wNFCI)} \\
\midrule
All & 1.038 & -1.385 & 0.087 & 0.199 \\
Selected & 1.020 & -1.451 & 0.077 & 0.210\\
Extracted & 1.030 & -1.463 & 0.075 & 0.198 \\
Filtered & 0.947 & -1.466 & 0.075 & 0.269 \\
Final Sample & 0.889 & -1.452 & 0.077 & 0.314 \\
\bottomrule
\end{tabular}
\end{table}

\clearpage

\subsection*{Figures}

\begin{figure}[htbp]
    \centering
    \includegraphics[width=0.9\textwidth]{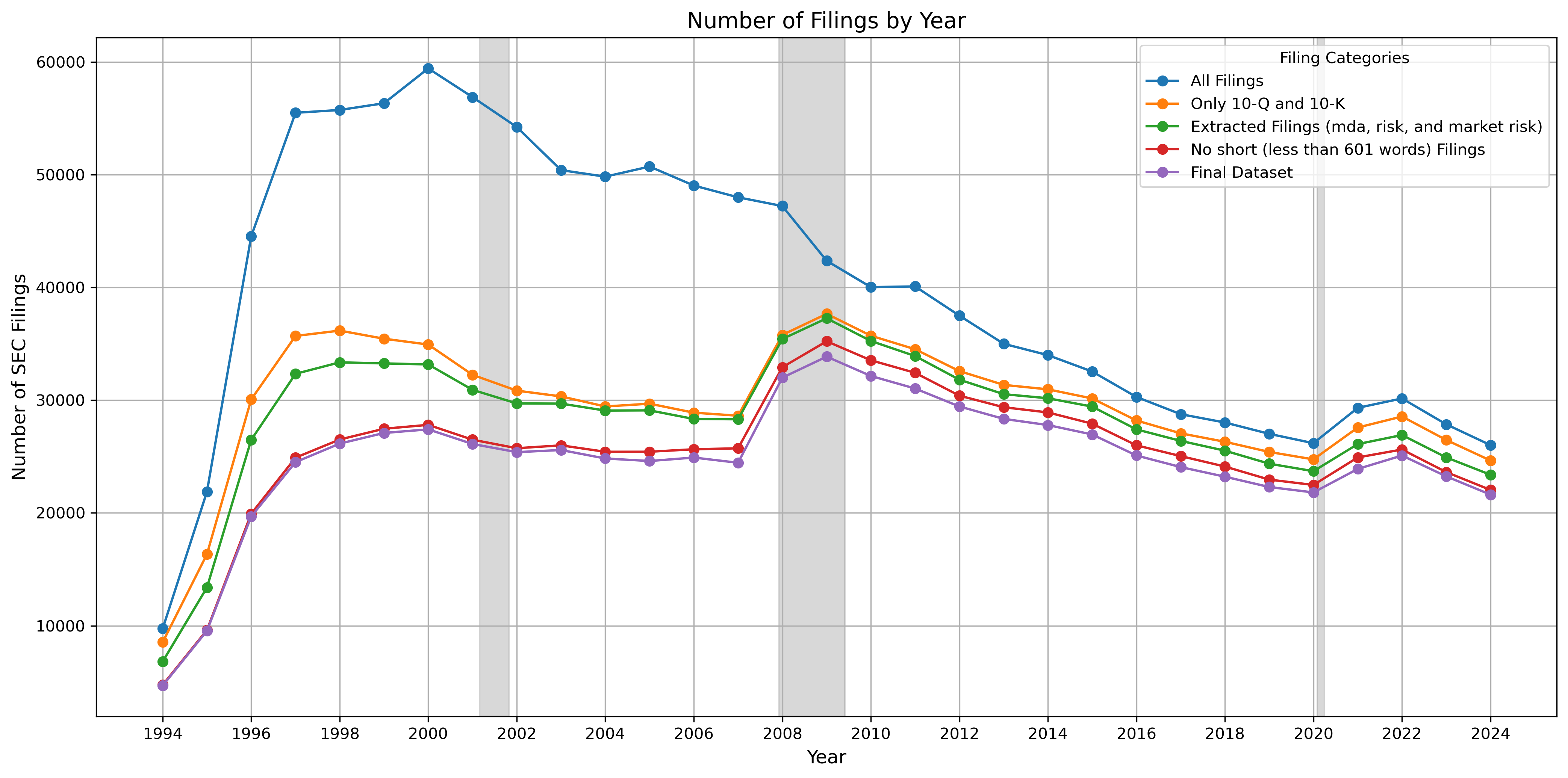}
    \caption{Number of Filings by Year}
    \label{fig:filings_by_year}
\end{figure}

\begin{figure}[htbp]
    \centering
    \includegraphics[width=0.9\textwidth]{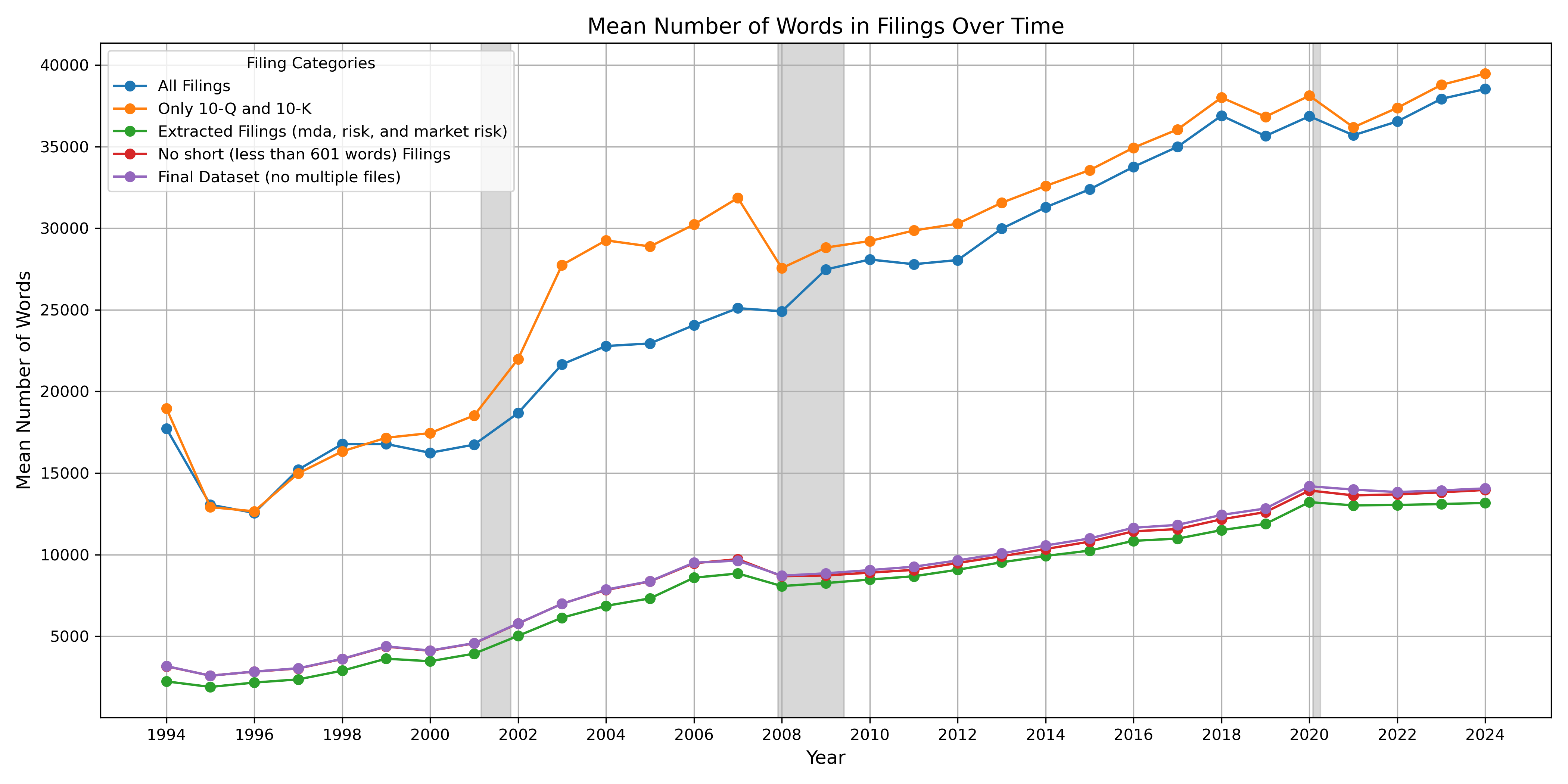}
    \caption{Mean Number of Words in Filings Over Time}
    \label{fig:mean_words_time}
\end{figure}

\begin{figure}[htbp]
    \centering
    \includegraphics[width=0.9\textwidth]{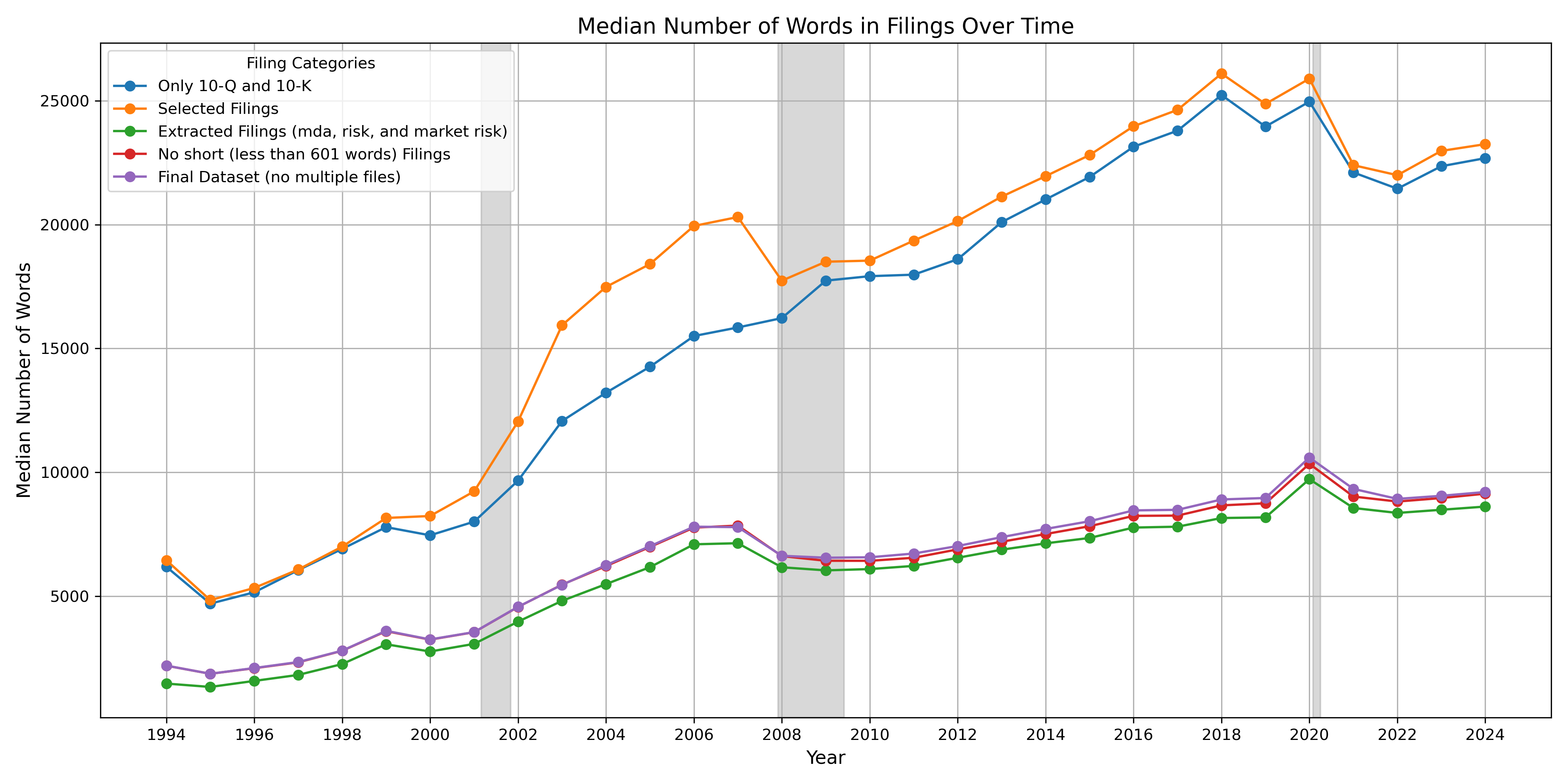}
    \caption{Median Number of Words in Filings Over Time}
    \label{fig:median_words_time}
\end{figure}

\begin{figure}[htbp]
    \centering
    \includegraphics[width=0.9\textwidth]{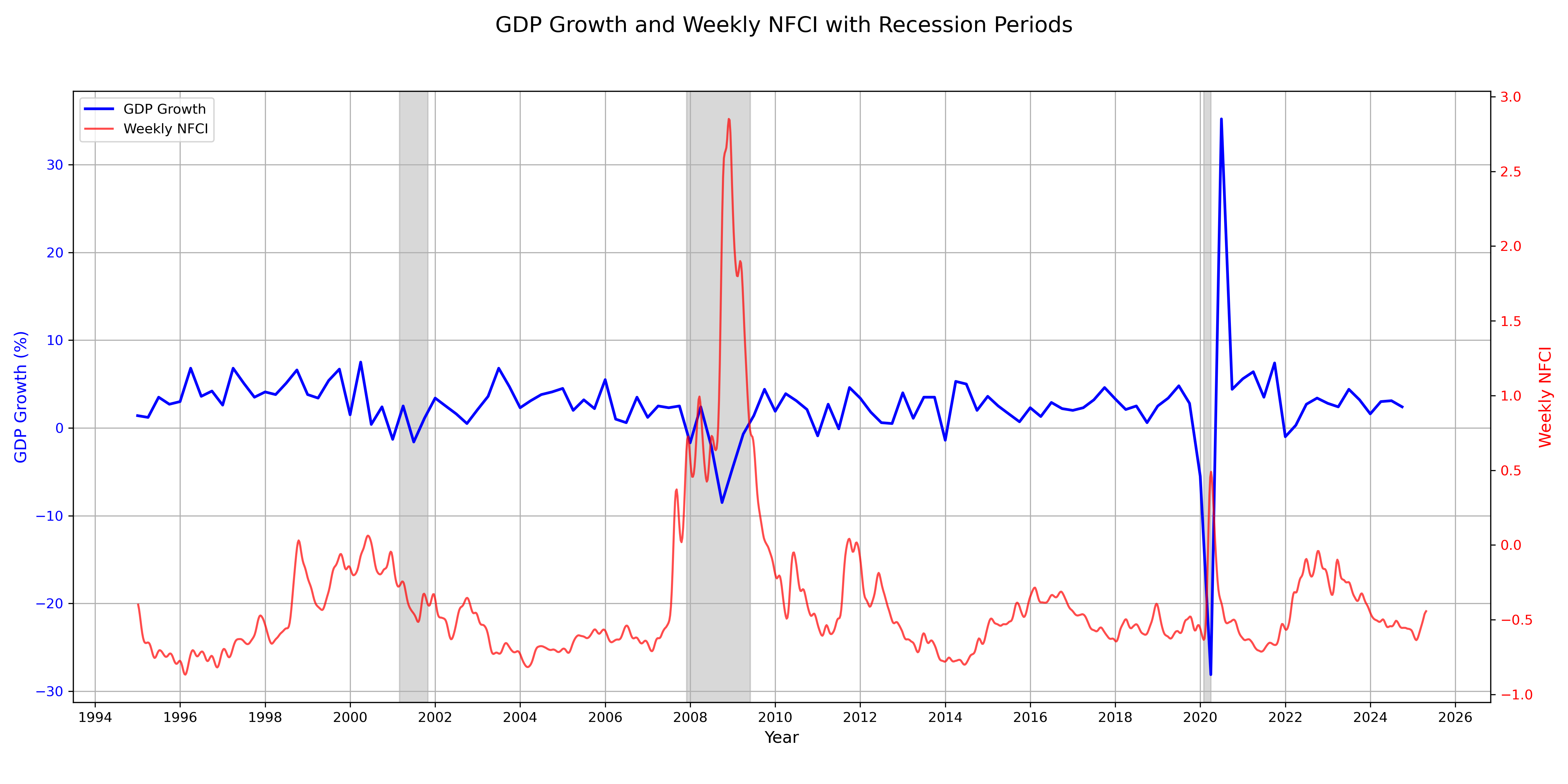}
    \caption{GDP Growth and Weekly NFCI Over Time}
    \label{fig:gdp_nfc_time}
\end{figure}

\begin{figure}[htbp]
    \centering
    \includegraphics[width=0.9\textwidth]{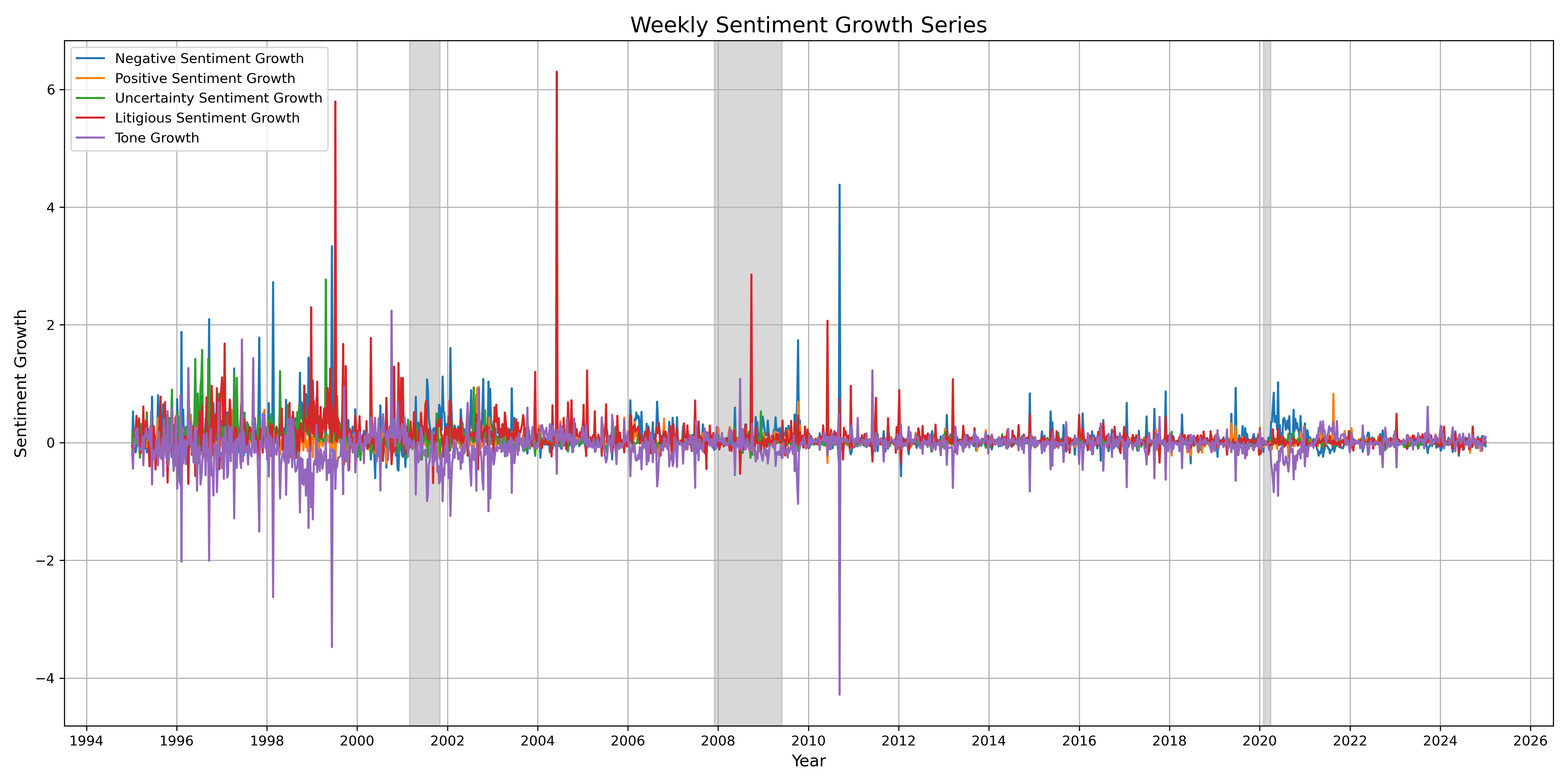}
    \caption{All Weekly Sentiment Growth Series}
    \label{fig:sentiment_series}
\end{figure}

\begin{figure}[htbp]
    \centering
    \includegraphics[width=0.9\textwidth]{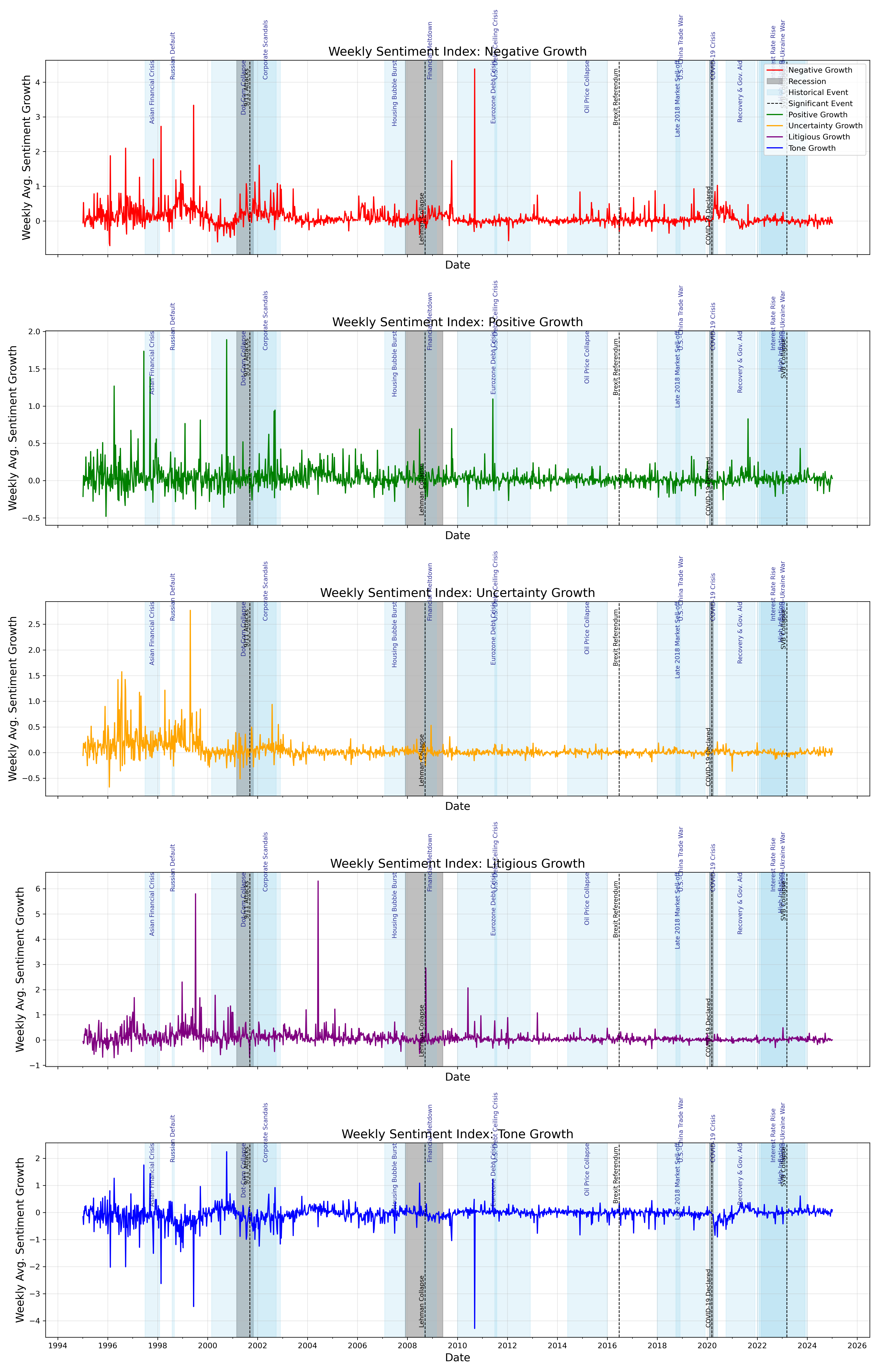}
    \caption{Weekly Sentiment Growth Series with Historical Events}
    \label{fig:sentiment_events}
\end{figure}

\begin{figure}[htbp]
\centering
\includegraphics[width=0.8\textwidth]{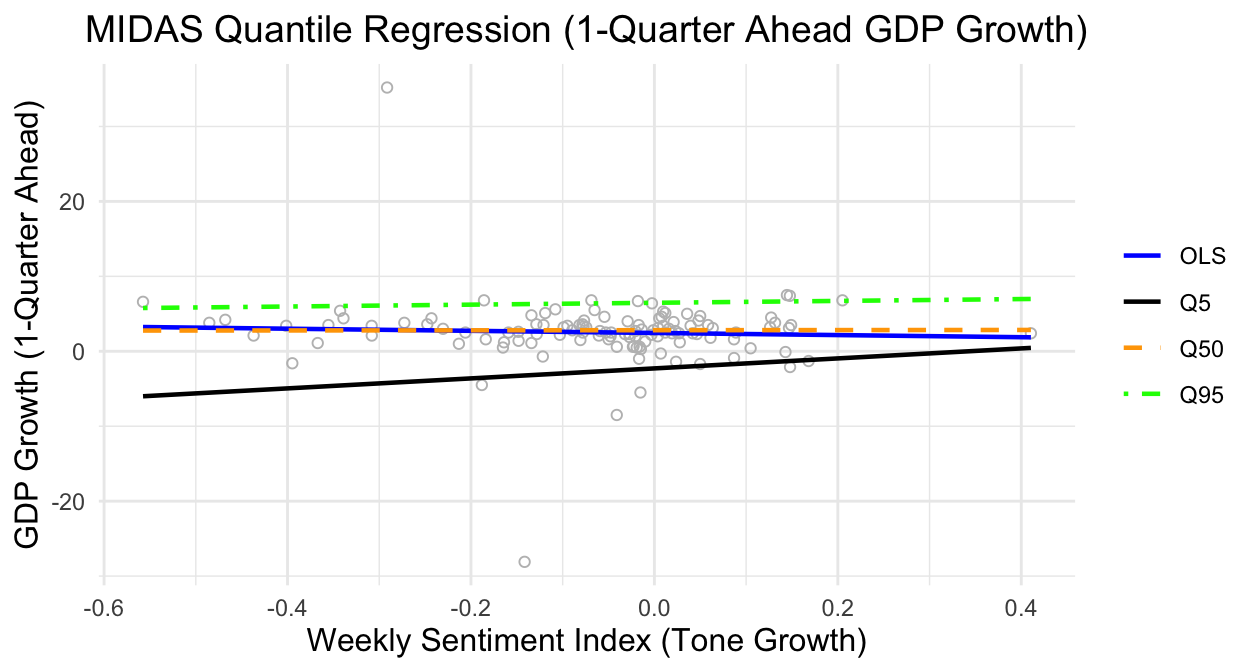}
\caption{One-quarter-ahead GDP growth vs. Weekly Sentiment Index (Tone Growth)}
\label{fig:gdp_vs_sentiment_scatter}
\end{figure} 

\begin{figure}[htbp]
\centering
\includegraphics[width=0.8\textwidth]{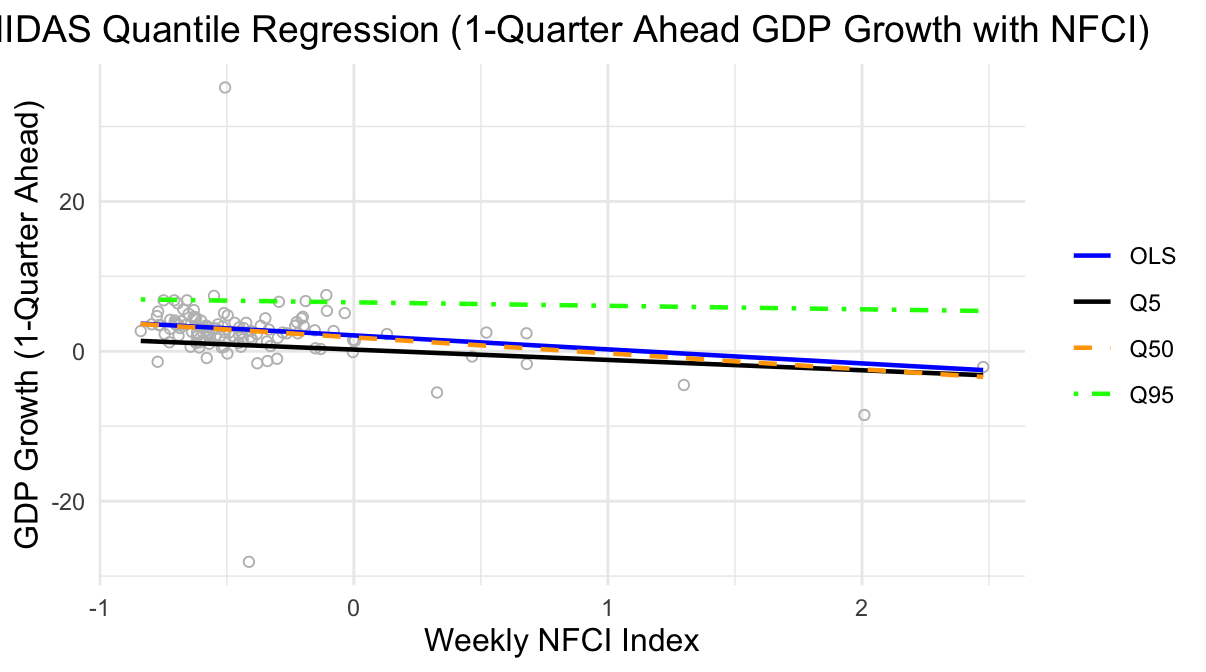}
\caption{One-quarter-ahead GDP growth vs. Weekly NFCI}
\label{fig:gdp_vs_NFCI_scatter}
\end{figure} 

\begin{figure}[htbp]
\centering
\includegraphics[width=0.8\textwidth]{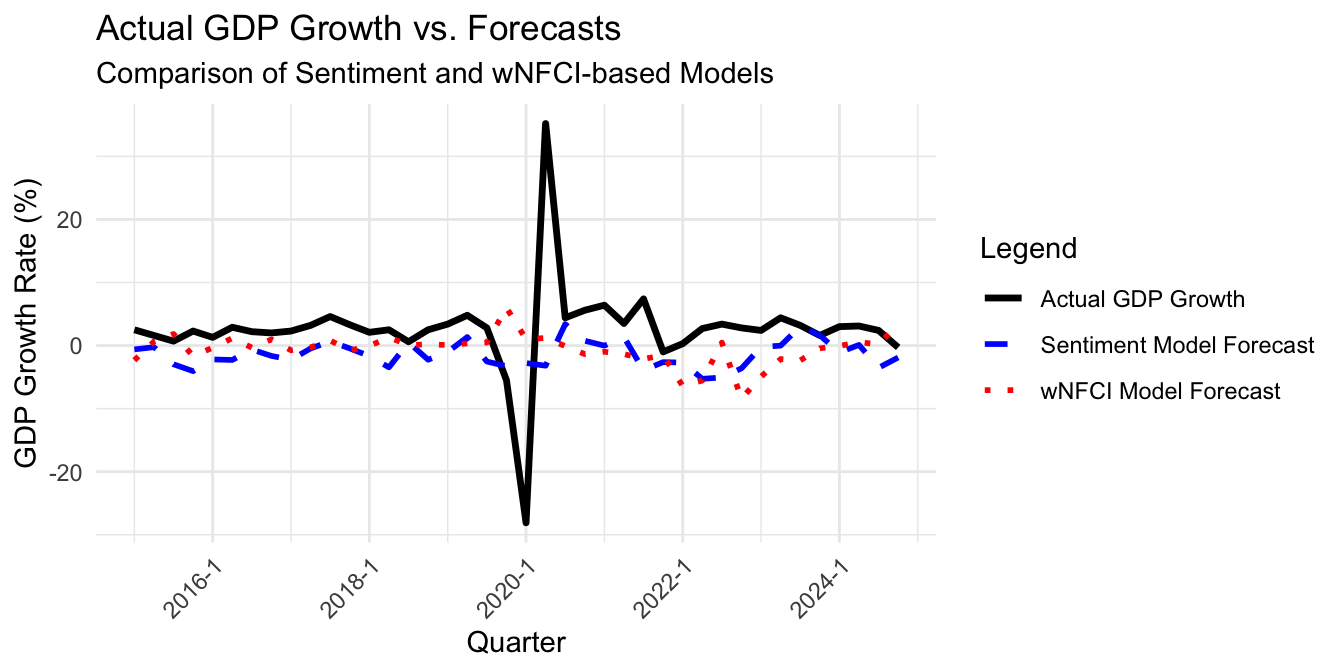}
\caption{Actual GDP Growth vs. Forecasts from Sentiment and wNFCI Models}
\label{fig:gdp_forecast}
\end{figure}

\begin{figure}[htbp]
\centering
\includegraphics[width=0.8\textwidth]{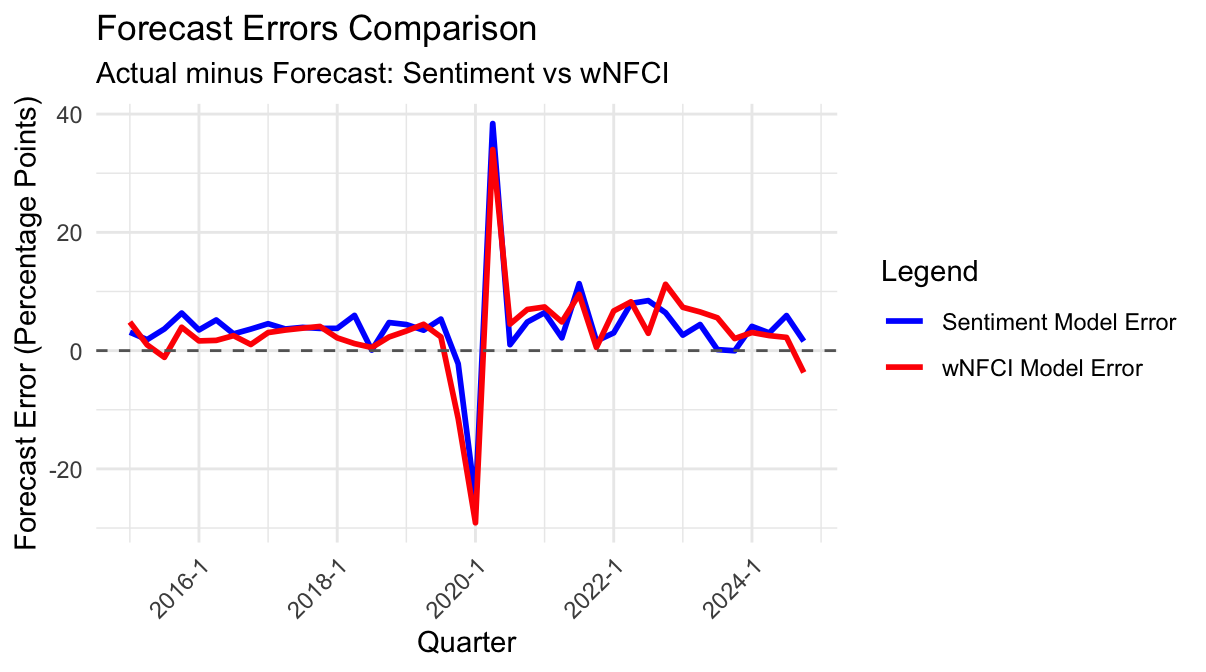}
\caption{Forecast Errors Comparison}
\label{fig:forecast_errors}
\end{figure}

\end{document}